\documentclass[aps,pre,reprint,showpacs,superscriptaddress]{revtex4-2}
\usepackage{chngcntr}
\usepackage{float}
\usepackage{color}
\usepackage{tabularx}
\usepackage{graphicx}
\usepackage{dcolumn}
\usepackage{bm}
\usepackage{mathrsfs}
\usepackage[ruled,vlined]{algorithm2e}
\usepackage{array} %
\usepackage{amssymb}
\usepackage[utf8]{inputenc}
\usepackage{tabstackengine}
\setstackEOL{\cr}
\usepackage{amsfonts,amsmath,mathrsfs,booktabs}
\usepackage{multirow,times}
\usepackage{float,color,bm}
\usepackage{empheq}
\usepackage{hhline}
\usepackage{tikz}
\usepackage{lipsum,mathtools}
\usepackage{diagbox}
\usepackage{bbm}
\usepackage[colorlinks,
linkcolor=blue,
anchorcolor=blue,
urlcolor=blue,
citecolor=blue
]{hyperref}
\newlength{\halfpagewidth}
\divide\halfpagewidth by 1

\newcommand{\splitatcommas}[1]{%
	\begingroup
	\ifnum\mathcode`,="8000
	\else
	\begingroup\lccode`~=`, \lowercase{\endgroup
		\edef~{\mathchar\the\mathcode`, \penalty0 \noexpand\hspace{0pt plus 0.3em}}%
	}\mathcode`,="8000
	\fi
	#1%
	\endgroup
}
\newcommand{\tuple}[1]{(\splitatcommas{#1})}
\newcommand{\set}[1]{\{\splitatcommas{#1}\}}
\newcommand{\RNum}[1]{\uppercase\expandafter{\romannumeral #1\relax}} 

\newcommand{\PreserveBackslash}[1]{\let\temp=\\#1\let\\=\temp}
\newcolumntype{C}[1]{>{\PreserveBackslash\centering}p{#1}}
\newcolumntype{R}[1]{>{\PreserveBackslash\raggedleft}p{#1}}
\newcolumntype{L}[1]{>{\PreserveBackslash\raggedright}p{#1}}
 \normalsize

\allowdisplaybreaks[3]
\begin{document}
	\title{Cooperation in Public Goods Games: Leveraging Other-Regarding Reinforcement Learning on Hypergraphs}
	\author{Bo-Ying Li}
	\address{School of Physics, Ningxia University, Yinchuan, 750021, P. R. China}
	\author{Zhen-Na Zhang}
	\address{School of Physics, Ningxia University, Yinchuan, 750021, P. R. China}
	\author{Guo-Zhong Zheng}
	\address{School of Physics and Information Technology, Shaanxi Normal University, Xi'an, 710062, P. R. China}
	\author{Chao-Ran Cai}
	\address{School of Physics, Northwest University, Xi'an, 710127, P. R. China}
	\author{Ji-Qiang Zhang}\email{zhangjq13@lzu.edu.cn }
	\address{School of Physics, Ningxia University, Yinchuan, 750021, P. R. China}
	\author{Chen Li}\email{chenl@snnu.edu.cn}
	\address{School of Physics and Information Technology, Shaanxi Normal University, Xi'an, 710062, P. R. China}
	
	\date{\today}
\begin{abstract}
Cooperation as a self-organized collective behavior plays a significant role in the evolution of ecosystems and human society. Reinforcement learning (RL) offers a new perspective, distinct from imitation learning in evolutionary games, for exploring the mechanisms underlying its emergence. However, most existing studies with the public good game (PGG) employ a self-regarding setup or are on pairwise interaction networks. Players in the real world, however, optimize their policies based not only on their histories but also on the histories of their co-players, and the game is played in a group manner.
In the work, we investigate the evolution of cooperation in the PGG under the other-regarding reinforcement learning evolutionary game (OR-RLEG) on hypergraph by combining the Q-learning algorithm and evolutionary game framework, where other players' action history is incorporated and the game is played on hypergraphs. Our results show that as the synergy factor $\hat{r}$ increases, the parameter interval is divided into three distinct regions — the absence of cooperation (AC), medium cooperation (MC), and high cooperation (HC) — accompanied by two abrupt transitions in the cooperation level near $\hat{r}_1^{*}$ and $\hat{r}_2^{*}$, respectively. Interestingly, we identify regular and anti-coordinated chessboard structures in the spatial pattern that positively contribute to the first cooperation transition but adversely affect the second. Furthermore,we provide a theoretical treatment for the first transition with an approximated $\hat{r}_1^{*}$, and reveal that players with a long-sighted perspective and low exploration rate are more likely to reciprocate kindness with each other, thus facilitating the emergence of cooperation. Our findings contribute to understanding the evolution of human cooperation, where other-regarding information and group interactions are commonplace.
\end{abstract}
\maketitle
\section{Introduction}\label{introduction}
Cooperation is both ubiquitous and significant in the evolution of human society and biological systems~\cite{griffin2004cooperation,van2002introduction,bernasconi1999cooperation,milinski2006stabilizing}, from altruistic pathogenic bacteria and ant fortress associations~\cite{bernasconi1999cooperation,d2018ecology} to community activities and civic participation in human society~\cite{van2002introduction,milinski2006stabilizing}. Deciphering the underlying mechanisms of how cooperative behavior evolved is crucial for the development of society~\cite{kennedy2005don}. The challenge focuses on why cooperation can be established and sustained in scenarios where there is a conflict between self-interest and group interests, such as the tragedy of commons~\cite{hardin1968tragedy}.

The evolutionary game (EG) theory, introduced and developed over the past fifty years, primarily aims to study the mechanisms behind the emergence of collective behaviors in ecosystems through natural selection~\cite{d2018ecology,smith1982evolution,ghoul2016ecology}. By viewing biological inheritance as akin to imitation learning (IL) in society, researchers have employed this framework to investigate self-organized, collective behaviors within human societies, such as cooperation~\cite{challet1997emergence,chen2015competition,traulsen2009exploration,sachs2004evolution}, fairness~\cite{debove2016models,rand2013evolution,zheng2023pinning}, epidemic dynamics~\cite{rogalski2017human,amaral2021epidemiological}, as well as urban eco-evolutionary processes~\cite{alberti2020complexity}. For the emergence of cooperation, the prisoner's dilemma game (PDG) model is often used, which illustrates the conflict between individual and collective interests in pairwise interactions~\cite{sachs2004evolution,doebeli2005models,rapoport1965prisoner,wang2022levy}. To extend this interaction to scenarios involving multiple players, researchers introduced the public goods game (PGG), which accommodates an arbitrary number of players~\cite{chen2015competition,rand2011evolution,helbing2010punish}. In PGGs, the conflict of interests between individuals and groups often leads to free-riding behaviors, which, if no countermeasure is taken, ultimately results in full defection~\cite{hardin1968tragedy,rand2011evolution,fischbacher2001people,fehr2000cooperation}. 
As a paradigmatic model, PGG has been widely employed to discuss many issues in society, such as resource management~\cite{wang2018replicator,meloni2017heterogeneous}, environmental protection~\cite{milinski2006stabilizing,tavoni2011inequality}, and epidemic prevention and control~\cite{reluga2010game,carande2007public}.  

After decades of effort, scholars have discovered various mechanisms that promote the emergence of cooperation in PGGs through both experimental and theoretical methods. In experiments, they have found that social reputation~\cite{milinski2006stabilizing}, information sharing~\cite{tavoni2011inequality}, social diversity~\cite{kurzban2001individual,cadsby1998gender}, dynamical reciprocity~\cite{liang2022dynamical}, and team competition~\cite{tan2007team} can enhance cooperation. From a theoretical perspective, mechanisms such as punishing free-riders and rewarding contributors~\cite{helbing2010punish,sigmund2001reward,szolnoki2010reward}, fostering competition~\cite{chen2015competition,szolnoki2011competition}, network heterogeneity~\cite{santos2008social}, reputation~\cite{hauert2010replicator}, and introducing noise~\cite{szolnoki2009topology} also contribute to suppressing the prevalence of free-riding behavior. These works provide significant insights into the mechanisms underlying the emergence of cooperation. 
However, most findings are based on IL with traditional pairwise interaction networks being assumed. The limitations are obvious:  IL primarily focuses on observing others, neglecting individuals’ experiences in human society, and pairwise networks fail to capture the group interactions inherent in PGGs.

The advancement of reinforcement learning (RL) and hypergraph theory offers the opportunity to address the two limitations. RL enables agents to gather direct experiences from their environment and adjust their behaviors based on received rewards, aiming to maximize cumulative gains \cite{sutton2018reinforcement,franccois2018introduction}. Hypergraph allows interactions across the entire group, rather than just pairwise interactions~\cite{battiston2020networks,bianconi2021higher}. In the hypergraph, the researcher found that the cooperation is enhanced by the heterogeneous investment~\cite{pan2023heterogeneous} and higher-order structures~\cite{burgio2020evolution} in the evolutionary game of PGGs. With the marriage of RL and EGs, some new insights into the mechanism of cooperation emergence for pairwise games have been provided, but primarily based on the PDG with two players~\cite{ding2023emergence,song2022reinforcement,jia2021local,yang2024interaction}. Only recently have a few studies begun to explore the evolution of multiplayer games within the RL framework using PGGs~\cite{jia2022empty,wang2023synergistic,zhang2024exploring,shen2024learning,zheng2024evolution}.
Ref.~\cite{jia2022empty} studied the evolution of cooperation in the PGG using traditional networks with empty nodes and the Bush-Mosteller model, finding that an appropriate proportion of empty nodes can stimulate individual cooperative behavior. Wang \emph{et al}. incorporated self-regarding Q-learning into evolutionary games to investigate the synergistic effects of learning rules and adaptive rewards on cooperation systematically~\cite{wang2023synergistic}. Furthermore, Ref.~\cite{zhang2024exploring} introduced loners into voluntary PGGs and revealed that the number of contributors becomes increasingly consistent under a large synergy factor and adjustments to the multiplier for smaller loner payoffs. By further integrating both IL and RL, the researchers discovered a unique semi-stable checkerboard pattern formed by defectors and cooperators~\cite{shen2024learning}. More recently, Ref.~\cite{zheng2024evolution} explored the evolution of cooperation in both Public Goods Games (PGGs) and voluntary PGGs using the Q-learning algorithm, leveraging environmental information. Compared to IL, their findings suggest cooperation is more likely to arise in RL operating through a distinct mechanism.

However, these existing studies primarily focus on either self-regarding reinforcement learning (SR-RL) on traditional networks. While SR-RL simplifies the model, it fails to align with our real-life gaming experiences, where decisions are influenced not only by our past actions but also by environmental cues, such as the historical actions of co-players. Traditional networks, on the other hand, fail to capture the complex group interaction inherent in PGGs. Therefore, we are particularly interested in the following questions: \emph{Does the other-regarding reinforcement learning (OR-RL) on hypergraph result in a distinct format of cooperation? If so, what are the underlying mechanisms behind this phenomenon?} Addressing these questions is crucial for understanding the emergence of cooperation in group interactions within human society from a reinforcement learning framework.

This paper is structured as follows: In Sec.~\ref{sec:model}, we introduce our other-regarding reinforcement learning evolutionary game (OR-RLEG) in PGGs on von Neumann hypergraphs, adopting the Q-learning algorithm.
In Sec.~\ref{sec:simulation}, We demonstrate that the average cooperation level, as a function of the synergy factor, is divided into three distinct regions by two transition points, each exhibiting notable differences in the spatial patterns of local cooperation levels within or between these regions. In Sec.~\ref{subsec:first_transition_point}, we determine the first transition point, which dictates whether cooperation emerges, using an analytic method based on further simulation. In Sec.~\ref{subsec:chessbord}, We focus on a distinctive form of cooperation—a chessboard structure emerging from temporal accumulation in spatial patterns—from the perspectives of local cooperation preferences and state transition modes. Our conclusions and discussions are presented in Sec.~\ref{sec:summary}.

\section{Model}\label{sec:model}

\begin{figure}[htbp]
	\centering
	\includegraphics[width=0.85\linewidth]{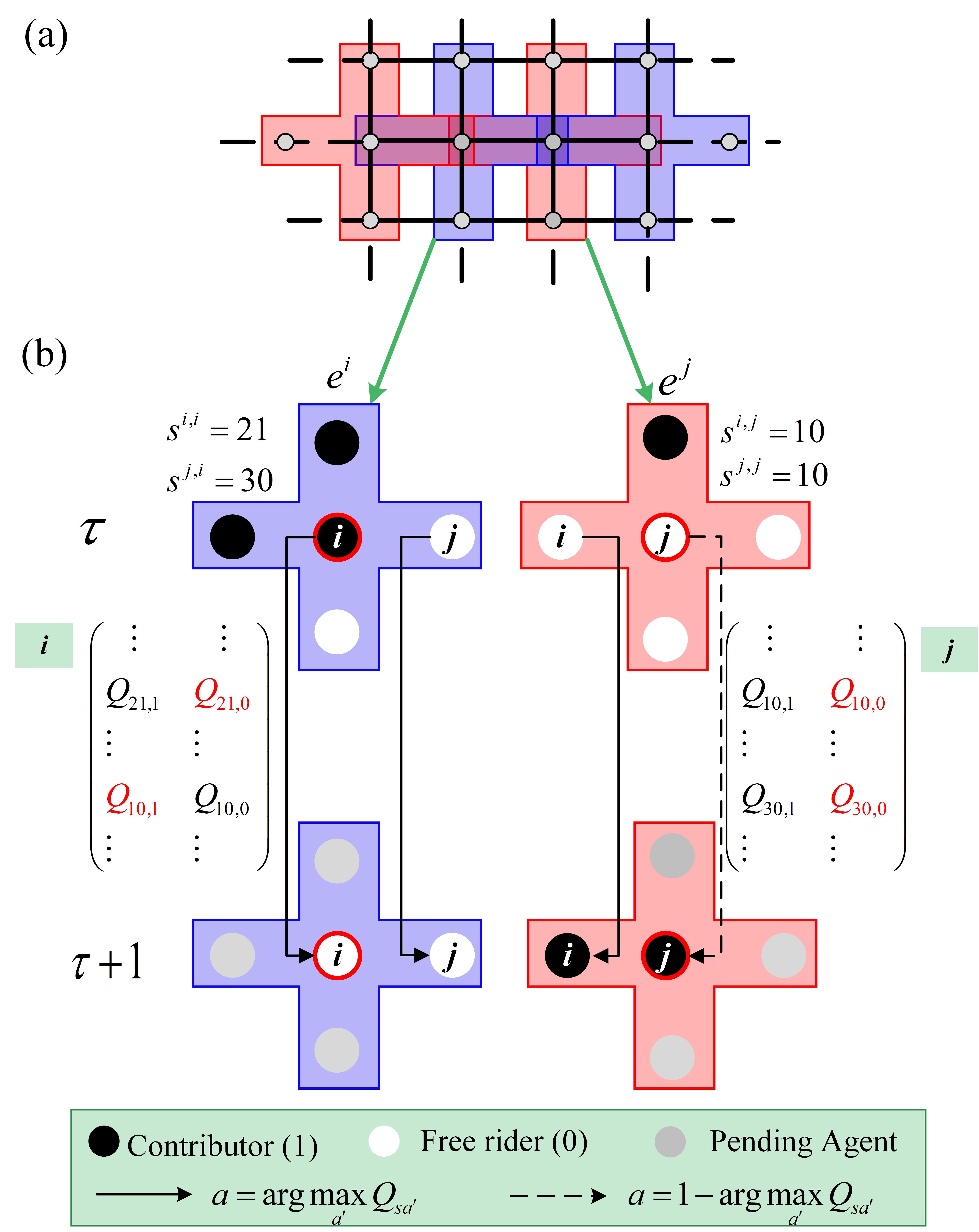}
	\caption{ (Color online) {\bf The hypergraph generated from von Neumann network and state-action schematic with Q-learning.} The hypergraph generated from the von Neumann network in our model is illustrated in (a). The hyperedges are color-coded, with each one encompassing $5$ agents represented by gray dots. To illustrate the state-action schematic with Q-learning in this hypergraph, the states of two agents $i$ and $j$, within the hyperedges $e^{i}$ and $e^{j}$ at the $\tau$th step, are shown in the top panel of (b). In each hyperedge, the initiator is marked with a red circular edge. Agents $i$ and $j$ select actions based on their states in $e^{i}$ and $e^{j}$ at the $\tau$th step and according to the policy outlined in Eq.~(\ref{eq:policy}) within their Q-tables. In the Q-tables, the maximum Q-value for each state is highlighted in red. The bottom panel of (b) shows the actions taken by $i$ and $j$ which partially determine their next states at $\tau + 1$. The exploitation-action and exploration-action are represented by solid and dashed lines, respectively.
	}\label{fig:hypergraph_map}
\end{figure}

In this study, we first introduce our other-regarding reinforcement learning evolutionary game (OR-RLEG) model within the framework of public goods games (PGGs), applied to a hypergraph $\mathcal{G} = \set{\mathcal{N},\mathcal{E}}$ generated from the von Neumann lattice. Initially, we configure $|\mathcal{N}|$ agents on an $L \times L$ von Neumann lattice with a periodic boundary. Following this, each agent $i\in \mathcal{N}$ along with its nearest neighbors in the von Neumann lattice, defines a set of nodes that forms a hyperedge $e^i\in \mathcal{E}$, as illustrated in Fig.~\ref{fig:hypergraph_map} (a). The hyperdegree of each agent is  $k = 5$, which is equal to the order of its hyperedge, denoted as $|e|$. The hypergraph is abbreviated as a von Neumann hypergraph.

At any step $\tau$ of the evolutionary dynamics, the update protocol
is divided into two processes: \emph{gaming} and \emph{learning} processes. During the \emph{gaming process}, each agent $i\in\mathcal{N}$ serving as an initiator will initiate a PGG and play with all the other agents (participants) in the hyperedge $e^{i}$. In the game, the initiator $i$ and any participants $j\in e^i$ select their actions $a^{i,i}(\tau)$ and $a^{j,i}(\tau)$ from the action set $\mathcal{A} = \{1, 0\}$, where $1$ and $0$ denote cooperation and defection, respectively. The agents that cooperate are referred to as ``contributors'', while those that defect are called ``free-riders''.
\begin{table}[htbp!]
	\centering
	\caption{{\bf The state set for agents.} In the table, $n_c$ denotes the number of contributors excluding the focal agent, and $a$ represents the agent’s action, both in the last step. The current state $s$ is determined by the action information in the last step.}
	\label{tab:states}
	\centering
	\begin{tabular}{|c|cc|}
		\hline
		\diagbox{$n_{c}(\tau-1)$}{$s(\tau)$}{$a(\tau-1)$} & $0$ & $1$ \\
		\hline
		$0$  &$s_{\tilde{0}}/00$ & $s_{0}/01$\\
		$1$  &$s_{\tilde{1}}/10$ & $s_{1}/11$\\
		$\vdots$ & $\vdots$ & $\vdots$ \\
		$4$  &$s_{\tilde{4}}/40$ & $s_{4}/41$ \\
		\hline
	\end{tabular}
\end{table}

During decision-making, agents use an RL algorithm known as Q-learning, which relies on their state information from the previous game round and learned experiences to seek optimal actions. However, unlike the self-regarding reinforcement learning evolutionary games (SR-RLEGs) in previous works~\cite{yang2024interaction,shen2024learning,zhang2020oscillatory,zou2024incorporating}, the state in our model is other-regarding. The state for any agent $j \in e^i$ can be expressed as 	$s^{j,i}(\tau) = n^{j,i}_{c}(\tau -1)a^{j,i}(\tau - 1)$, where $n^{j,i}_c(\tau-1) = \sum_{k\in e^{i}_{-j}} a^{k,i}(\tau-1)$ is the number of contributors other than $j$ and $a^{j,i}(\tau - 1)$ is
$j$'s action in the previous step.
Then, the state set for agents is $\mathcal{S} = \set{s_0, s_1, \cdots, s_{4}, s_{\tilde{0}}, s_{\tilde{1}}, \cdots, s_{\tilde{4}}} = \set{01, 11, \cdots, 41, 00, 10, \cdots, 40}$ (see Tab.~\ref{tab:states}). To simplify the notations, we employ $s_{m}$ and $s_{\tilde{m}}$ here to differentiate a contributor and a free-rider, while both have $m$ other contributors within the hyperedge $e^i$ at $\tau  - 1$. While experiential cognition is characterized by the agent's Q-table, where each element $Q_{s,a}(\tau)$ as a map of $\mathcal{S}\times\mathcal{A}\rightarrow\mathbb{R}$, represents the action value for $a$ at $s$ in cognition (see Tab.~\ref{tab:Qtable}). 
\begin{table}[htbp!]	
	\centering
	\caption{{\bf The Q-table for agents.} In the table, $Q_{sa}$ denotes the action value for column-action $a$ within the row-state $s$.}
	\label{tab:Qtable}
	\begin{tabular}{|c|cc|}
		\hline
		\diagbox{$s$}{$Q_{sa}$}{$a$} & $0$ & $1$ \\
		\hline
		$s_{0}$  & $Q_{s_{0},0}$ & $Q_{s_{0},1}$  \\
		$\vdots$  & $\vdots$ & $\vdots$       \\
		$s_{4}$ & $Q_{s_{4},0}$ & $Q_{s_{4},1}$                      \\
		$s_{\tilde{0}}$ & $Q_{s_{\tilde{0}},0}$ & $Q_{s_{\tilde{0}},1}$ \\      
		$\vdots$   & $\vdots$  &  $\vdots$  \\
		$s_{\tilde{4}}$ & $Q_{s_{\tilde{4}},0}$ & $Q_{s_{\tilde{1}},1}$ \\  
		\hline
	\end{tabular}
\end{table}

Then, each agent $j \in e^{i}$ takes action based on its own Q-table and state $s^{j,i}$ at 
$\tau$. The policy for action selection is as follows
\begin{eqnarray}\label{eq:policy}
	a^{j,i}(\tau)&=&\pi\left(s^{j,i}(\tau), {\bf Q}^{j}(\tau)\right) 
	\nonumber\\
	&=&\left\{
	\begin{array}{lc} 
		1-\arg\max\limits_{a'} Q_{s^{j,i},a'}(\tau), & \frac{\epsilon}{|\mathcal{A}|}, \\
		\arg\max\limits_{a'} Q_{s^{j,i},a'}(\tau), & 1 - \epsilon + \frac{\epsilon}{|\mathcal{A}|}.
	\end{array}
	\right.
\end{eqnarray}
In other words, $j$ selects $1 - \arg\max_{a'} Q_{s^{j,i},a'}(\tau)$ as its action, referred to as the \emph{exploration-action}, with the probability $\frac{\epsilon}{|\mathcal{A}|}$, and otherwise chooses $\arg\max_{a'} Q_{s^{j,i},a'}(\tau)$ as its action, referred to as the \emph{exploitation-action}. Here, the exploration rate, $0<\epsilon\ll 1$, serves to adjust the trade-off between exploration and exploitation. 
The state-action schematic is shown in Fig.~\ref{fig:hypergraph_map}.

After the decision-making, the initiator $i$ within the hyperedge $e^{i}$ receives its payoff at $\tau$, which is determined based on the actions taken by the agents within $e^i$ as follows
\begin{eqnarray}\label{eq:payoff}
	\Pi^{i}(\tau) = \frac{r}{|e^i|}\cdot\sum\limits_{j \in e^i} a^{j,i}(\tau) - a^{i,i}(\tau).
\end{eqnarray}	
Here, $r/|e^i|= \hat{r}$, serving as the game parameter, represents the synergy factor that defines the benefit of mutual cooperation among agents within $e^i$.


During the learning process, only the initiator $i$ within $e^i$ updates the action-value $Q_{s,a}$ based on new experiences, while the participants do not. The only difference between the initiator and participants reflects that each initiator is actively engaged in the game, seeking higher rewards and continuously elevating its cognition through the Q-table. In contrast, the participants are only passively involved in the proposed games, without the expectation of enhancing their cognition. Therefore, the update for $Q^{i}_{s,a}$ only utilizes state $s = s^{i,i}$ and actions $a = a^{i,i}$ related to $\tau$ and $\tau + 1$ steps. 
To simplify, the following expression omits the hyperedge information and agent identifiers,
\begin{eqnarray}\label{eq:Q_update}
	Q^{i}_{s,a}(\tau+1) &=& (1-\alpha)\cdot Q^{i}_{s,a}(\tau) + \alpha \times \nonumber\\
	&\quad&\left[ \Pi^{i}(\tau) + \gamma\cdot\max\limits_{a^{\prime}} Q^{i}_{s^{\prime},a^{\prime}}(\tau) \right].
\end{eqnarray}	
Here, $\alpha\in (0,1]$ represents the learning rate, indicating the degree to which new experiences override old ones. While $\gamma\in [0,1)$ is the discounting factor that determines the importance of future rewards since $\max_{a'} Q^{i}_{s',a'}$ is the maximum action value that $i$ can expect within the next state $s^{\prime} = s^{i,i}(\tau+1)$. 
While at the end of the step, any agent $j \in e^{i}$ will update its state according to actions of the agents in $e^{i}$ as follows 
\begin{eqnarray}\label{eq:update_state}
	s^{j,i}(\tau+1) = n^{j,i}_{c}(\tau)a^{j,i}(\tau).
\end{eqnarray}
Note that each agent, whether initiated the game or participated, has only one Q-table available to handle action selection based on its state in the corresponding hyperedge.
\begin{algorithm}[htbp!]  
	\caption{The Algorithm of OR-RLEG in the Context of PGG}\label{algorithm:protocol}
	\LinesNumbered 
	\KwIn{Learning parameters: $\alpha$, $\gamma$, $\epsilon$; Population:$\mathcal{N}$}
	{\bf Initialization}\;
	\For{$i$ in $\mathcal{N}$}{
		Create a Q-table with each item in the matrix near zero\;
		\For{$j$ in $e^i$}{ 
			Pick an action $a^{j,i}$ randomly from $\mathcal{A}$\;
			Generate state:$s^{j,i} \rightarrow n^{j,i}_{c}a^{j,i}$\;
		}	
	}
	\Repeat{\text the system becomes statistically stable or evolves for the desired time duration}
	{{\bf Gaming process}\;
		\For{$i$ in $\mathcal{N}$}{
			Initiate a public goods game for any agent in $e^i$\;
			\For{$j$ in $e^i$}{
				Generate the action $a^{j,i}$ according to $s^{j,i}$, ${\bf Q}^{j}$ and Eq.~(\ref{eq:policy})\;	
			}
			Get reward $\Pi^{i}$ according to Eq.~(\ref{eq:payoff})\; 
		}
		{{\bf Learning process}\;
			\For{$i$ in $\mathcal{N}$}{	
				Update Q-table according to Eq.~(\ref{eq:Q_update})\;
				\For{$j$ in $e^i$}{
					Update state $s^{j,i}$ according to Eq.~(\ref{eq:update_state});	
				}
			}
		}
	}
\end{algorithm}

In simulation, the two processes repeat until the system becomes statistically stable or evolves for the desired time duration. To summarize, the pseudo-code is provided in Algorithm~\ref{algorithm:protocol}. For easy reference, we display descriptions of the mathematical notation used in the models, as detailed in Tab.~\ref{tab:notation1} in Sec.~\ref{sec:notations}. By default in this work, the combination of learning parameters is $\tuple{\alpha, \gamma, \epsilon} = \tuple{0.1, 0.9, 0.01}$ for OR-RLEGs unless otherwise specified. The network used is the von Neumann hypergraph. 

To measure the local cooperation level within any hyperedge $e^{i}$
at $\tau$, we define $f_c^{i}(\tau)$ as the number of contributors in $e^i$,
\begin{equation}\label{eq:fci}
	f_c^{i}(\tau) = {\sum\limits_{j \in e^i} a^{j,i}(\tau)}/{|e^i|},
\end{equation} 
Then, the global cooperation level in all hyperedges at $\tau$ is 
\begin{eqnarray}\label{eq:fc}
	f_{c}(\tau) :=\sum_{e^{i}\in\mathcal{E}}f^{i}_{c}(\tau)/|\mathcal{E}|,
\end{eqnarray}
and average cooperation level $f_{c}(\tau)$ in global across the stable stage is
\begin{eqnarray}\label{eq:bar_fc}
	\bar{f}_{c}:=\sum_{\tau = t_0}^{t} f_{c}(\tau)/(t-t_0).  
\end{eqnarray}
Here, $t_0$ represents any specific step when the system has reached stability. 
In addition, we also focus on the agent's cooperation preference within all games that it is involved in, that defined as
\begin{eqnarray}\label{eq:pci}
	p_c^{i}(\tau) := \sum\limits_{e^{j} \in \{e|i\in e\}} a^{i,j}(\tau)/{k^i}.
\end{eqnarray}
Here, $\{\cdots\}$ denotes the set that includes all hyperedges containing the agent $i$.
	
\section{Simulation Results}\label{sec:simulation}

\begin{figure}[htbp]
	\centering
	\includegraphics[width=0.85\linewidth]{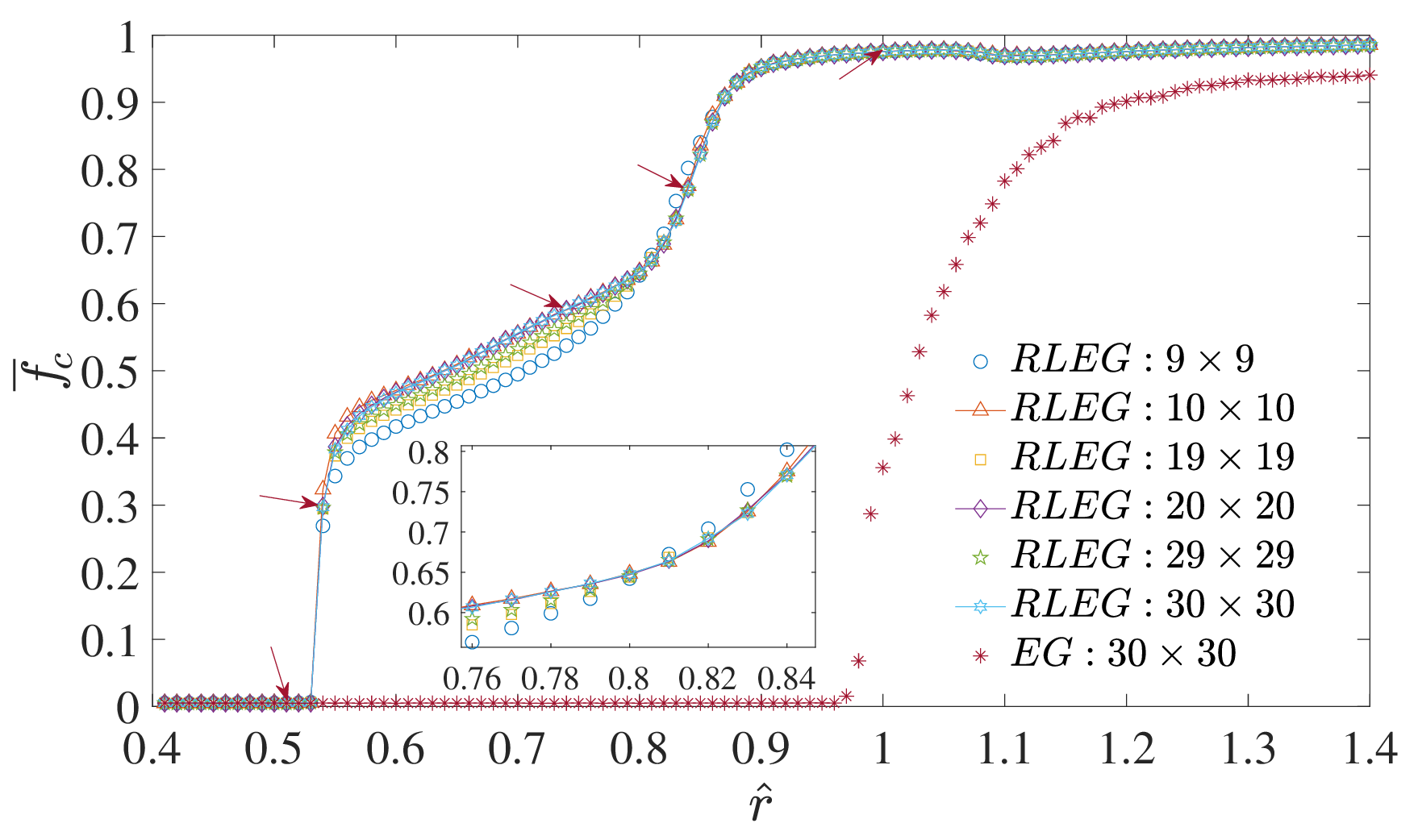}
	\caption{ (Color online) {\bf The global average cooperation level as a function of the game parameter.} The global average cooperation level $\bar{f}_{c}$ as a function of synergy factor $\hat{r}$ in OR-RLEGs, across various population scales $|\mathcal{N}| = L\times L$ on hypergraph, is shown to investigate how $|\mathcal{N}|$ affect $\bar{f}_{c}$. The default learning parameters are $\tuple{\alpha, \gamma, \epsilon} = \tuple{0.1, 0.9, 0.01}$ in OR-RLRGs. The counting consists of $10^6$ steps after $10^{10}$ transient steps. For comparison, we also display $\bar{f}_{c}$ as the function of $\hat{r}$ in evolutionary games on hypergraphs. Here, the selection intensity and mutation probability are $\beta = 10$ and $\mu = 0.01$ under Fermi learning rule [see Sec.~\ref{sec:app_EG_model}].     
	}
	\label{fig:global_fc}
\end{figure}

In Fig.~\ref{fig:global_fc}, we primarily exhibit the average global cooperation level $\bar{f}_{c}$ as the function of the game parameter $\hat{r}$ in our OR-RLEG model. As a benchmark, we also include a comparison with traditional evolutionary games (EGs). Similar to EGs, the results exhibit that $\bar{f}_{c}$ in our model increases with synergy factor $\hat{r}$ when $\hat{r}$ is greater than a transition point $\hat{r}_{1}^*$. However, $\bar{f}_{c}$ in the model consistently exceeds that in EGs as long as $\hat{r} > \hat{r}_{1}^{*}$. In addition, $\hat{r}_{1}^* \approx 0.534$ for OR-RLEGs, which is much lower than $\hat{r}^{*}\approx 0.96$ for EGs.
Furthermore, the results highlight another distinct difference between the two models: $\bar{f}_{c}$ as the function of $\hat{r}$ in our model be delineated into three distinct regions - \emph{absence of cooperation} (AC), \emph{medium cooperation} (MC), and \emph{high cooperation} (HC). There is a sharp increase not only at  $\hat{r}_{1}^{*}$, the transition point from the AC to the MC region, but also between the MC and HC region. This indicates there may be another transition point $\hat{r}_{2}^*$ between MC and HC. 

To approximately identify the location of $\hat{r}_{2}^*$, we examine how the system size $|\mathcal{N}|$ affects the transition points based on the phase transition theory~\cite{privman1990finite}. The result shows $\bar{f}_c$ exhibits a distinct crossing point at $\hat{r} = 0.803$ when comparing odd- and even-sized systems. In addition, for a given $\hat{r}$ near the point, $\bar{f}_c$ remains essentially constant with increasing $|\mathcal{N}|$ in the case of even-sized $|\mathcal{N}|$. However, for the odd-sized $|\mathcal{N}|$, $\bar{f}_c$ is increased with $|\mathcal{N}|$ when $\hat{r}$ is to the left of the point, while it decreases with an increase of $|\mathcal{N}|$ when $\hat{r}$ is to the right of the point. Therefore, it is reasonable to conjecture that the crossing point corresponds to the second transition point $\hat{r}^{*}_{2}$.

\begin{figure}[htbp]
	\centering
	\includegraphics[width=\linewidth]{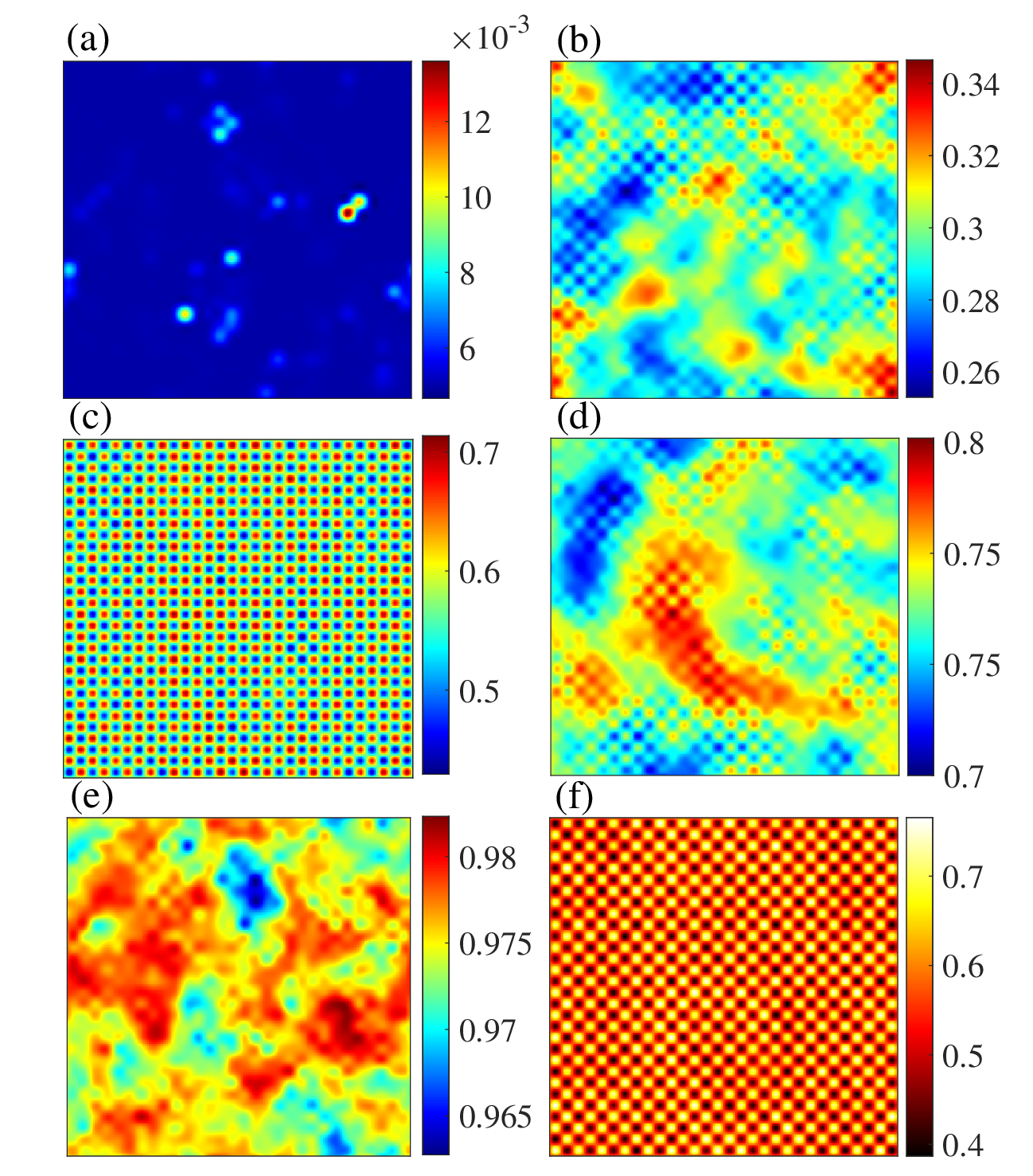}
	\caption{(Color online) {\bf Spatial patterns for averaged cooperation levels locally and a spatial pattern for average cooperation preferences of agents.} Panels (a - e) display the pattern of average local cooperation levels over time, denoted as $\bar{f}_c^{i}$, across the hypergraph for various game parameters [see the red arrows in Fig.~\ref{fig:global_fc} (a)]: in the absence of cooperation (AC), between AC and medium cooperation (MC), in MC, between MC and high cooperation (HC), and in HC regions, respectively. The game parameters in (a - e) are $\hat{r} = 0.50, 0.54, 0.72, 0.84$, and $1.00$, respectively. In (f), we show the pattern of average cooperation preference of agents over time, denoted as $\bar{p}_{c}^{i}$, using the same game parameters as in (c). Under this condition, both $\bar{f}_c^{i}$ and $\bar{p}_c^{i}$ exhibit a chessboard pattern with alternating strengths and weaknesses, but are negatively correlated with each other. In (a - f), the scale of population size is $|\mathcal{N}| = 30\times 30$. The average for $f_{c}^{i}(\tau)$ and $p_{c}^{i}(\tau)$ are computed over $10^8$ steps, after $10^{10}$ transient steps.}
	\label{fig:local_fc}
\end{figure}

To investigate $\bar{f}_{c}$ from a local perspective, we examine the averaged local cooperation level $\bar{f}^{i}_{c}$ within $e^{i}\in\mathcal{E}$ over $10^6$ steps during the stable stage, and illustrate it through spatial patterns in Fig.~\ref{fig:local_fc}. Within the AC, the results show that $\bar{f}^{i}_{c}$ is only slightly higher within a few hyperedges compared to the others, as shown in (a). While in the area between the AC and MC regions, (b) shows different clusters emerge, and the average cooperation level in each cluster differs from that of its neighboring clusters. Furthermore, distinctly different from previous works in EGs, each cluster here is composed of blurry \emph{chessboard structures} rather than a homogeneous one, characterized by a regular alternation of high and low $\bar{f}^{i}_{c}$.  

As $\hat{r}$ enters the MC region, the clusters that appeared in the area between the AC and MC merge into a single dominant one, and the composed chessboard structure for it becomes clear [see Fig.~\ref{fig:local_fc} (c)]. In (d), the result displays $\bar{f}^{i}_{c}$ in the area between MC and HC will further increase. However, similar to the area between the AC and HC regions, clusters reemerge, and the chessboard structure within these clusters becomes unclear once again. As $\hat{r}$ enters the HC region, we observe that the chessboard structure almost disappears, and the local cooperation level in each hyperedge becomes very high.

To further investigate the cooperation preference for each agent, Fig.~\ref{fig:local_fc} (f) displays the average cooperation preference $\bar{p}^{i}_{c}$ in spatial pattern as $\hat{r}$ is within MC region. Much like the spatial pattern observed for $\bar{f}^{i}_{c}$, the pattern for $\bar{p}^{i}_{c}$ also exhibits distinct \emph{chessboard structures}. This indicates that agents who prefer to serve as contributors and those who prefer to be free-riders are alternately arranged in space. But, $\bar{p}^{i}_{c}$ is negatively correlated  with $\bar{f}^{i}_{c}$ in spatial pattern. In other words, if $\bar{f}_{c}^{i}$ within the hyperedge $e^{i}$ is higher than that within the neighboring hyperedges, then agent $i$'s cooperation preference $\bar{p}_{c}^{i}$ across all games will be lower than that of its neighbors, and vice versa [see (f) and (c)]. This suggests that the \emph{anti-coordination} seen in Snowdrift Games also appears in the context of PGGs in OR-RLEGs, but in the form of \emph{partial anti-coordination} instead of complete anti-coordination.

Figure~\ref{fig:local_fc} also uncovers why the global level of cooperation $\bar{f}_c$ is influenced by the population scale when $|\mathcal{N}|$ is odd, but not when $|\mathcal{N}|$ is even [see Fig.~\ref{fig:global_fc}]. Clearly, an odd-sized $|\mathcal{N}|$ will hinder the system from forming a large chessboard pattern, particularly at a global level. Specifically, the smaller odd-sized $|\mathcal{N}|$, the more detrimental it is to the formation of such pattern [see Fig.~\ref{fig:app_pattern2} (a, b)]. 
This is the reason that $\bar{f}_c$ decreases with $|\mathcal{N}|$ when $|\mathcal{N}|$ is odd in the MC region, but increases as $|\mathcal{N}|$ decreases in the HC region. Following this clue, we can infer that the anti-coordinated chessboard structure will enhance cooperation in global, when $\hat{r}^*_1<\hat{r}<\hat{r}^*_2$, but suppress it when $\hat{r}>\hat{r}^*_2$.

\begin{figure}[htbp]
	\centering
	\includegraphics[width=\linewidth]{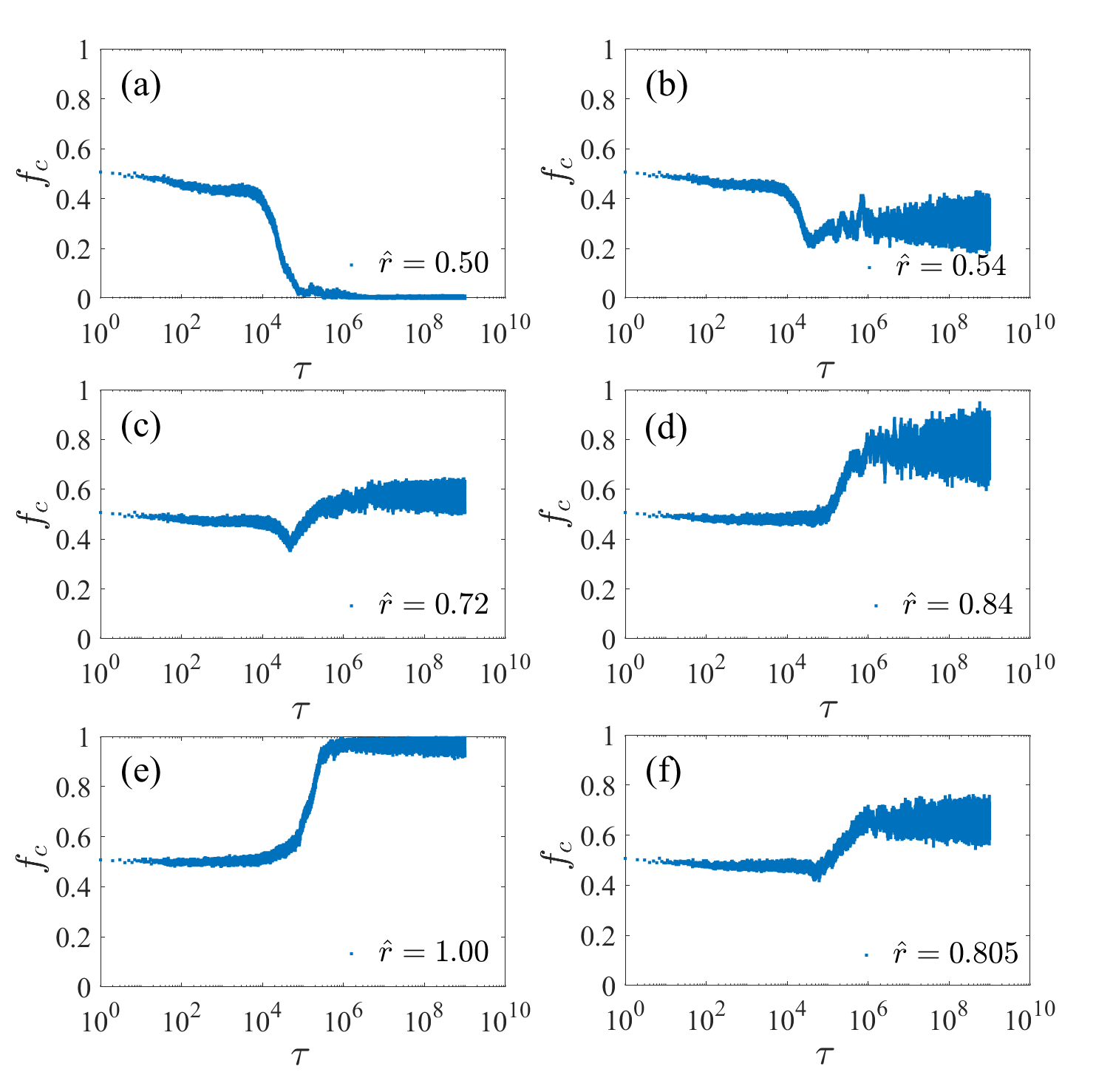}
	\caption{(Color online) {\bf The time series of global cooperation level for different game parameters.} Panels (a - f) exhibit the time series of global cooperation levels, $f_c(\tau)$, under different game parameters: in the AC, between AC and MC, in MC, between MC and HC, in HC, and at the second transition point $\hat{r}^*_{2}$, respectively. The population size is $|\mathcal{N}| = 30\times 30$, and other parameters are the same as Fig.~\ref{fig:local_fc}.}
	\label{fig:time_series}
\end{figure}

Furthermore, to examine global cooperation level through evolving dynamics, we present the time series of $f_{c}$ in Fig.~\ref{fig:time_series}. Similar to Fig.~\ref{fig:local_fc}, we also select one $\hat{r}$ in each of the different regions and additionally include a specific one, which is at the second transition point $\hat{r}^*_2$. For $\hat{r}$ within the AC region, Fig.~\ref{fig:time_series} (a) reveals that $f_{c}$ initially decreases slowly, then experiences a rapid decline after $\tau\approx 10^4$, and finally approaches $0$. In contrast, for $\hat{r}$ in between AC and MC regions or MC region, increases rapidly after the rapid decline and stabilizes at a certain value [see (b) and (c)]. When $\hat{r}$ falls within between MC and HC regions or the HC region, Fig.~\ref{fig:time_series} (d) and (e) show that $f_{c}$ initially experiences a slight decrease, followed by a rapid increase, and ultimately stabilizes. For $\hat{r}$ at the transition point $\hat{r}^*_2$, (f) shows $f_{c}$ that in the area between the MC and HC regions, where the rapid decrease of $f_{c}$ almost ceases. 

Based on the time series shown in Fig.~\ref{fig:time_series}, we conjecture that the first transition point $\hat{r}^*_1$ separates cases where $f_{c}$ either rebounds after a rapid decline or does not. And, the second one $\hat{r}^*_2$ distinguishes between cases where $f_{c}$ will undergo a rapid decline or not.

\section{Mechanism Analysis}\label{sec:analysis}
\subsection{Transition Point for the Emergence of Cooperation}\label{subsec:first_transition_point}
\subsubsection{Further Simulations for Analysis}\label{subsubsec:simulation_for_analysis}
\begin{figure}[htbp]
	\centering
	\includegraphics[width=0.9\linewidth]{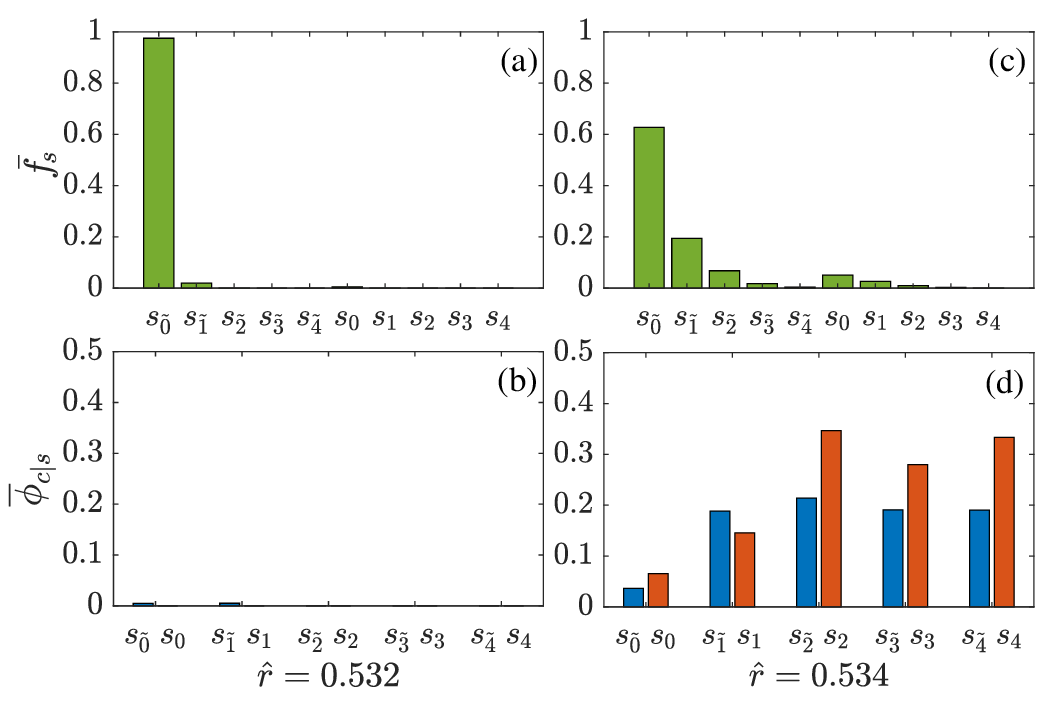}
	\caption{ (Color online) {\bf The distribution of different states, the cooperation preferences for various given states near the first transition point.} Panels (a) and (b) show $\bar{f_{s}}$ and $\bar{\phi}_{c|s}$ for any $s\in\mathcal{S}$ as defined in Eqs.~(\ref{eq:fs1}) and ~(\ref{eq:phi1}) at $\hat{r}_{1}^{*-} = 0.532$, whereas (c) and (d) illustrate these quantities at $\hat{r}_{1}^{*+} = 0.534$. In (a - d), the population size is $|\mathcal{N}| = 30\times 30$, and other parameters are the same as Fig.~\ref{fig:local_fc}. 
	}
	\label{fig:modes}
\end{figure}  
To investigate the transition point $\hat{r}_{1}^{*}$ in the emergence of cooperation, we first analyze the distribution of states for initiators and their cooperation preferences across all states, especially around $\hat{r}_{1}^{*}$. The probability that initiators in population are in state $s\in \mathcal{S}$ is defined as follows 
\begin{eqnarray}\label{eq:fs1}
	\bar{f}_s := \frac{\sum\limits_{\tau = t_0}^{t}\sum\limits_{i\in\mathcal{N}}\mathbbm{1}_{s^{i,i}(\tau) = s}}{|\mathcal{N}|(t-t_{0})},
\end{eqnarray}
which meets normalization $\sum_{s\in\mathcal{S}} \bar{f}_s = 1$. 
Here, $\mathbbm{1}_{\emph{predicate}}$ denotes the random variable that is $1$ if $predicate$ is true and $0$ if it is not.
Based on the initiator's action in different states, we further define the cooperation preference of initiators in a given state $s\in \mathcal{S}$ as follows,
\begin{eqnarray}\label{eq:phi1}
	\bar{\phi}_{c|s} := \frac{\sum\limits_{\tau = t_0}^{t}\sum\limits_{i\in\mathcal{N}} \mathbbm{1}_{s^{i,i}(\tau) = s,a^{i,i}(\tau) = 1}}{\sum\limits_{\tau = t_0}^{t}\sum\limits_{i\in\mathcal{N}}\mathbbm{1}_{s^{i,i}(\tau) = s}}.
\end{eqnarray}   
Here, we appoint $\bar{\phi}_{c|s}$ to $0$ in a special situation where the denominator is $0$.

Finally, to determine the local state transition modes in the hyperedges, we also investigate Cohen’s kappa matrix $[\kappa(s, s^{\prime})]$. In this matrix, the element $\kappa(s, s^{\prime})$ represents Cohen’s kappa coefficient for the temporal correlations between consecutive states $s$ and $s^{\prime}$, as defined below
\begin{eqnarray}\label{eq:kappa1}
	\kappa(s,s^{\prime}):= \frac{\bar{f}(s,s^{\prime})-\bar{f}(s)\bar{f}(s^{\prime})}{1-\bar{f}(s)\bar{f}(s^{\prime})}.
\end{eqnarray}
Here,
\begin{eqnarray}
	\bar{f}(s,s^{\prime}):=  \frac{\sum\limits_{\tau=t_0}^{t-1}\sum\limits_{i\in\mathcal{N}}\mathbbm{1}_{s^{i,i}(\tau)=s,s^{i,i}(\tau+1)=s^{\prime}}}{|\mathcal{N}|(t-t_0-1)}. \nonumber
\end{eqnarray}	
For state transitions, a self-loop mode can be reflected by a positive diagonal element in the matrix, while a local cyclic mode with a period-2 is indicated by pairs of positive symmetric elements. Conversely, in the matrix, a positive $\kappa(s,s)$ indicates that the transition of $s$ forms a local self-looped $s\leftrightarrow s$ mode. Furthermore, a pair of positive $\kappa(s,s^{\prime})$ and $\kappa(s^{\prime},s)$, with $\kappa(s,s^{\prime})\approx \kappa(s^{\prime},s)$, suggest that the transition between $s$ and $s^{\prime}$ forms a cyclic $s\leftrightarrow s^{\prime}$ mode.   

For $\hat{r}\rightarrow \hat{r}_{1}^{*-}$, Fig.~\ref{fig:modes} (a) shows that the state $s_{\tilde{0}} = 00$ becomes dominant because an initiator will not cooperate at $\tau$ within a hyperedge, which is nearly independent of the state at the previous step, as illustrated in (b).
In contrast, for $\hat{r}\rightarrow \hat{r}_{1}^{*+}$, Fig.~\ref{fig:modes} (c) shows that there are new states, $s_{\tilde{1}} = 10$ and $s_{0} = 01$, emerge although $s_{\tilde{0}} = 00$ still dominates. 
This occurs because a free-rider in the current step will reciprocate the kindness shown by contributors in the next step. Specifically, an initiator will choose cooperation in the next step if one of the participants serves as a contributor in the current step [see (d)].
Since our focus is on how cooperation emerges from its absence as $\hat{r}$ increases, we compare the differences in $\bar{\phi}_{c|s}$ between these two scenarios with $s_{\tilde{0}} = 00$ and $s_{0} = 01$. The results suggest there is a balance of competition between the two-state transition modes at $\hat{r}_{1}^{*}$: self-looped $00\leftrightarrow 00$ mode and cyclic $10\leftrightarrow 01$ mode. 
\begin{figure}[htbp]
	\centering
	\begin{minipage}{0.5\textwidth}
		\centering
		\includegraphics[width=0.9\linewidth]{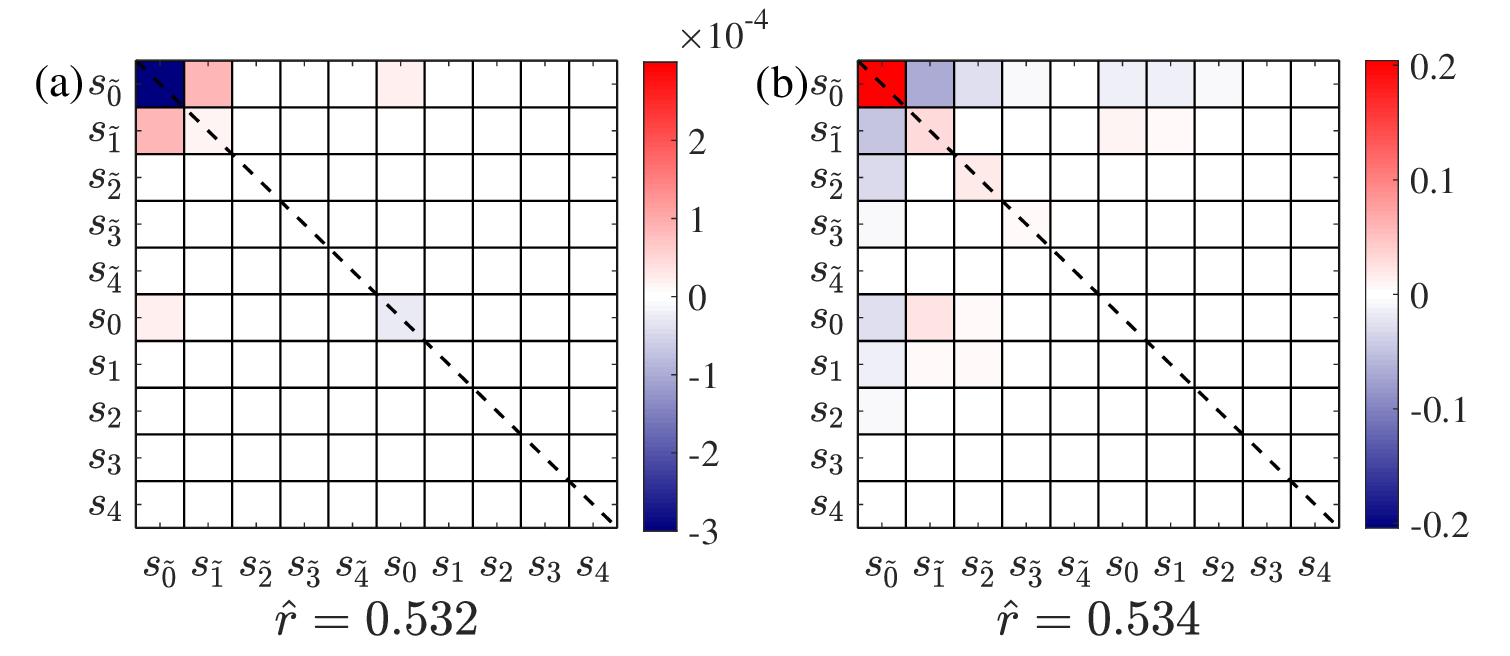}
		\includegraphics[width=0.85\linewidth]{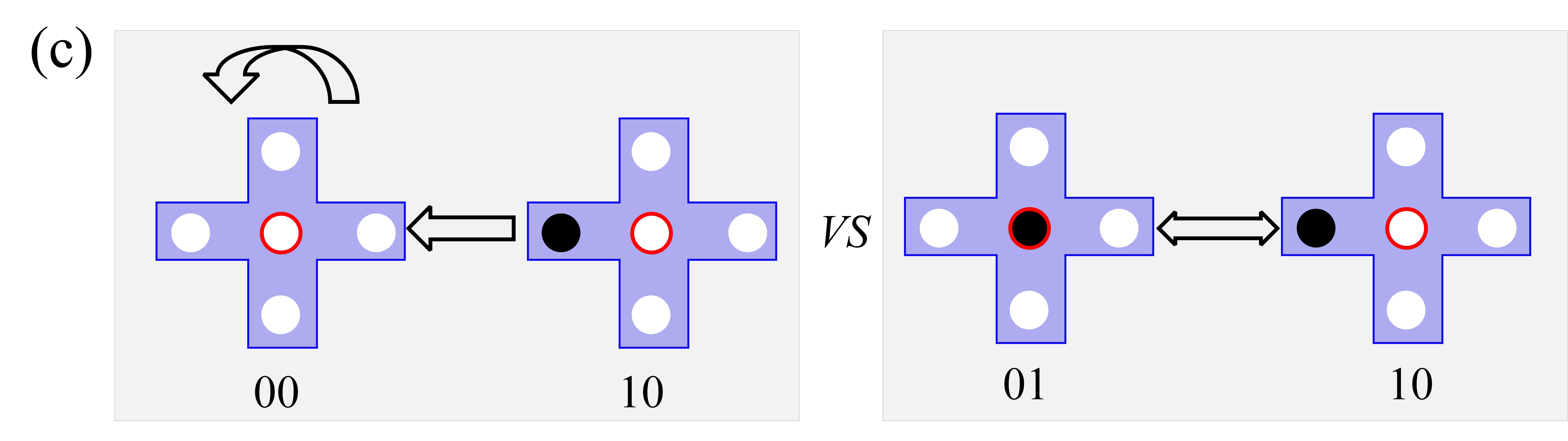}
	\end{minipage}	
	\caption{{\bf Cohen’s kappa coefficient matrices for temporal correlations near the first transition point and competing modes at the transition point.} 
		Panels (a) and (b) show the matrices of Cohen’s kappa coefficients for temporal correlations at $\hat{r}_{1}^{*-} = 0.532$ and $\hat{r}_{1}^{*+} = 0.534$. The element $\kappa(s, s^{\prime})$ in $[\kappa(s,s^{\prime})]$ is the coefficient of correlation between consecutive states $s$ and $s^{\prime}$. A positive diagonal element $\kappa(s,s)$ suggests the presence of a local self-loop $s\leftrightarrow s$ mode in the hyperedges, while a pair of positive, symmetric non-diagonal elements $\kappa(s,s^{\prime})$ and $\kappa(s^{\prime},s)$, with $\kappa(s,s^{\prime}) \approx \kappa(s^{\prime},s)$, indicates the existence of a local cyclic $s\leftrightarrow s^{\prime}$ mode. 
		(c) illustrates the competing modes at the first transition point $\hat{r}_{1}^{*} \approx 0.5332$ in simulation, which are the self-looped mode $00\leftrightarrow 00$ and the cyclic  mode $10\leftrightarrow 01$. The initiator in each hyperedge is marked with a red circular edge.	
		In (a) and (b), the averages are computed over $10^8$ steps, after $10^{10}$ transient steps.
	}\label{fig:kappa1}
\end{figure}

To verify the suggestion, we present the matrix of Cohen’s kappa coefficients for temporal correlations between consecutive states at $\hat{r}_{1}^{*-} = 0.532$ and $\hat{r}_{1}^{*+} = 0.534$, as shown in Fig.~\ref{fig:kappa1}. At $\hat{r}_{1}^{*-} = 0.532$, (a) demonstrates that the initiator’s state transitions from $00$ to $01$ or $10$ by either itself or the participant, and then quickly returns to state $00$ again based on the actions determined by the Q-tables. In contrast at $\hat{r}_{1}^{*+} = 0.534$, a new cyclic $01\leftrightarrow 10$ mode emerge after transitioning from $00$ to $01$ or $10$ [see (b)]. The results confirm the suggestion on competing modes from Fig.~\ref{fig:modes}. Thus, $00\leftrightarrow 00$ and $10\leftrightarrow 01$ modes exhibit competing action selections in the state $10$, still defection or cooperation [see (c)]. In other words, our key question about competing temporal modes is whether a free-rider will reciprocate the kindness of another contributor or continue to free-ride when encountering such kindness.

\subsubsection{Analysis of Transition Point via Competing Modes}\label{subsubsec:analysis_transion_point1}
Based on further simulations, our focus should be on the expected values for defection and cooperation in state $10$, specifically $Q_{10,0}$ in self-looped $00\leftrightarrow 00$ mode and  $Q_{10,1}$ in cyclic $10\leftrightarrow 01$ mode.
Here, we focus on the expectation of $Q_{10,0}$ in self-looped $00\leftrightarrow 00$ mode firstly. The evolutionary dynamics of $Q_{00,0}$ and $Q_{10,0}$ in $00\leftrightarrow 00$ mode under exploration as follows
\begin{subequations}\label{eq:00}
	\begin{align}
		Q_{00,0}\rightarrow& Q_{00,0}(1-\alpha)+\binom{|e|-1}{0}\left(\frac{\epsilon}{2}\right)^0\left(1-\frac{\epsilon}{2}\right)^{|e|-1} \nonumber\\
		&\times\alpha\left(\Pi_{00}+\gamma Q_{00,0}\right) 
		+\binom{|e|-1}{1}\frac{\epsilon}{2}\left(1-\frac{\epsilon}{2}\right)^{|e|-2} \nonumber\\
		&\times\alpha\left(\Pi_{10}+\gamma\max_{a'}Q_{10,a'}\right)+O(\epsilon^2), \label{eq:00_Q000}\\
		Q_{10,0}\rightarrow& Q_{10,0}(1-\alpha)+\binom{|e|-1}{0}\left(\frac{\epsilon}{2}\right)^0\left(1-\frac{\epsilon}{2}\right)^{|e|-1} \nonumber\\
		&\times\alpha\left(\Pi_{00}+\gamma Q_{00,0}\right) 
		+\binom{|e|-1}{1}\frac{\epsilon}{2}\left(1-\frac{\epsilon}{2}\right)^{|e|-2} \nonumber\\
		&\times\alpha\left(\Pi_{10}+\gamma\max_{a'}Q_{10,a'}\right)+O(\epsilon^2). \label{eq:00_Q100}
	\end{align}
\end{subequations}
Here, the second term on the right-hand side of the above equations represents how corresponding updates to the Q-table for the initiator affect the expectations of $Q_{00,0}$ and of $Q_{10,0}$, assuming no participant selects the exploration-actions in their corresponding state. And, the third term illustrates how these updates impact the expectations when only one of the participants chooses its exploration-action. All other cases, where more participants choose exploration-actions, are encompassed within $O(\epsilon^2)$. 
By omitting $O(\epsilon)$ further, the stable $Q_{00,0}^*$ and $Q_{10,0}^*$ in Eq.~\eqref{eq:00} meet
\begin{subequations}
	\begin{align}
		Q_{00,0}^* &\approx (1-\alpha)Q_{00,0}^*+\alpha\left(1-\frac{\epsilon}{2}\right)^{|e|-1}\left(\Pi_{00}+\gamma Q_{00,0}^{*}\right),  \label{eq:stable_00_Q000}\\
		Q_{10,0}^* &\approx (1-\alpha)Q_{10,0}^*+\alpha\left(1-\frac{\epsilon}{2}\right)^{|e|-1}\left(\Pi_{00}+\gamma Q_{00,0}^{*}\right).  \label{eq:stable_00_Q100}
	\end{align}
\end{subequations}
Then, we can get the stable expectation of $Q_{00,0}$ and expectation of $Q_{10,0}$ in $00\leftrightarrow 00$ mode as follows
\begin{eqnarray}
	Q_{00,0}^{*}=Q_{10,0}^{*}\approx\displaystyle{\frac{(1-\frac{\epsilon}{2})^{|e|-1}\Pi_{00}}{1-\gamma(1-\frac{\epsilon}{2})^{|e|-1}}}.  
\end{eqnarray}

Then, we pay attention to the expectation of $Q_{10,1}$ in the competing $01\leftrightarrow 10$ mode. In Eqs.~\eqref{eq:1001_Q101} and \eqref{eq:1001_Q010}, we give evolutionary dynamics of $Q_{10,1}$ and $Q_{01,0}$ in $01\leftrightarrow 10$ mode under exploration as follows
\begin{subequations}\label{eq:1001_Q}
	\begin{align}
		Q_{10,1}\rightarrow& \displaystyle{Q_{10,1}(1-\alpha)+\binom{|e|-1}{0}\left(\frac{\epsilon}{2}\right)^0(1-\frac{\epsilon}{2})^{|e|-1}} \nonumber\\
		&\times\displaystyle{\alpha\left(\Pi_{01}+\gamma Q_{01,0}\right)} 
		+\displaystyle{\binom{|e|-1}{1}\frac{\epsilon}{2}(1-\frac{\epsilon}{2})^{|e|-2}}\nonumber\\
		&\times\displaystyle{\alpha\left(\Pi_{11}+\gamma \max_{a^{\prime}}Q_{11,a^{\prime}}\right) + O(\epsilon^2)},  \label{eq:1001_Q101}\\
		Q_{01,0}\rightarrow& 
		\displaystyle{Q_{01,0}(1-\alpha)+\binom{|e|-1}{0}\left(\frac{\epsilon}{2}\right)^{0}(1-\frac{\epsilon}{2})^{|e|-1} }, \nonumber\\
		&\times\displaystyle{\alpha\left(\Pi_{10}+\gamma Q_{10,1}\right)}
		+\displaystyle{\binom{|e|-2}{1}\frac{\epsilon}{2}(1-\frac{\epsilon}{2})^{|e|-2}} \nonumber\\
		&\times\displaystyle{\alpha\left(\Pi_{20}+\gamma \max_{a^{\prime}}Q_{20,a^{\prime}}\right) } + \displaystyle{\frac{\epsilon}{2}(1-\frac{\epsilon}{2})^{|e|-2} } \nonumber\\ 
		&\times\displaystyle{\alpha \left(\Pi_{00}+\gamma\max_{a^{\prime}}Q_{00,a^{\prime}} \right) + O(\epsilon^2)}.  \label{eq:1001_Q010}
	\end{align}
\end{subequations}
Similar to Eqs.~\eqref{eq:00_Q000} and ~\eqref{eq:00_Q100}, the second term on the right-hand side of Eqs.~\eqref{eq:1001_Q101} and \eqref{eq:1001_Q010} show how corresponding updates to the Q-table for the initiator within this hyperedge respectively affect the expectations of $Q_{10,1}$ and of $Q_{01,0}$, assuming no participant chooses the exploration-action in their respective states. The rest terms except $O(\epsilon^2)$, on the other hand, show how these updates affect the expectations when only one of the participants takes exploration-action. By omitting $O(\epsilon)$ in Eq.~(\ref{eq:1001_Q}), we learn the stable $Q_{01,0}^*$ and $Q_{10,1}^*$ meet
\begin{eqnarray}
	\left(
	\begin{array}{c}
		Q_{10,1}^{*} \\
		Q_{01,0}^{*}
	\end{array}
	\right) &\approx&
	\left(
	\begin{array}{cc}
		1-\alpha & \alpha\gamma(1-\frac{\epsilon}{2})^{|e|-1} \\
		\alpha\gamma(1-\frac{\epsilon}{2})^{|e|-1} & 1-\alpha
	\end{array}
	\right)\cdot
	\nonumber\\
	&\quad&\left(
	\begin{array}{c}
		Q_{10,1}^{*} \\
		Q_{01,0}^{*}
	\end{array}
	\right) +
	\alpha(1-\frac{\epsilon}{2})^{|e|-1}
	\left(	
	\begin{array}{c}
		\Pi_{01} \\
		\Pi_{10}
	\end{array}
	\right).
\end{eqnarray}
Then, we can get $Q_{01,0}^*$ and $Q_{01,1}^*$ that are
\begin{subequations}
	\begin{align}
		Q_{10,1}^*&\approx \displaystyle{\frac{(1-\frac{\epsilon}{2})^{|e|}\left[\Pi_{10}(1-\frac{\epsilon}{2})+\gamma(1-\frac{\epsilon}{2})^{|e|}\Pi_{01}\right]}{(1-\frac{\epsilon}{2})^2-\gamma^2(1-\frac{\epsilon}{2})^{2|e|}} },\\
		Q_{01,0}^*& \approx\displaystyle{\frac{(1-\frac{\epsilon}{2})^{|e|}\left[\Pi_{01}(1-\frac{\epsilon}{2})+\gamma(1-\frac{\epsilon}{2})^{|e|}\Pi_{10}\right]}{(1-\frac{\epsilon}{2})^2-\gamma^2(1-\frac{\epsilon}{2})^{2|e|}}}.
	\end{align}
\end{subequations}

\begin{figure}[htbp]
	\centering
	\includegraphics[width=0.95\linewidth]{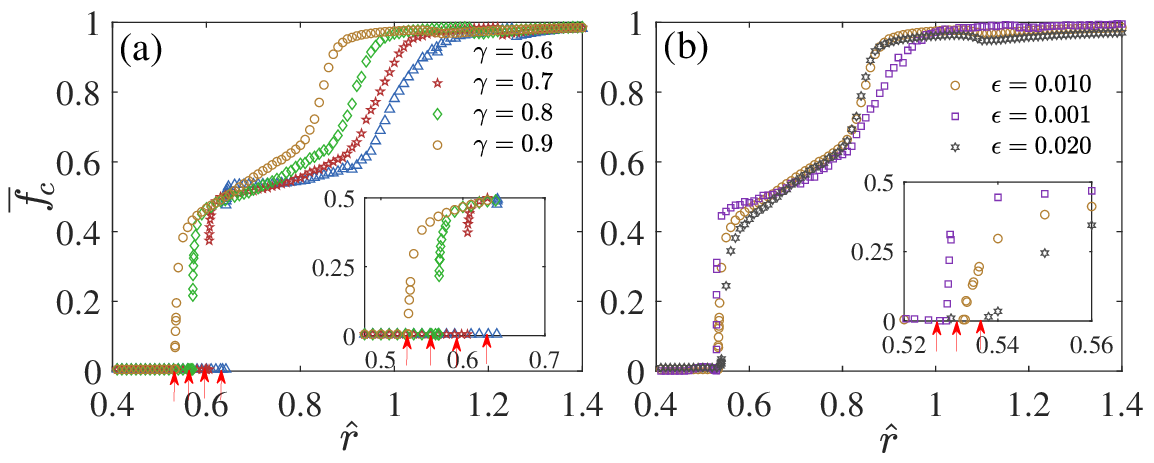}	
	\caption{ (Color online) {\bf The average global cooperation level as the function of the game parameter under different learning parameters.} Panel (a) shows the function $\bar{f}_{c}(\hat{r})$ under different values of discount factor $\gamma$, while (b) presents the function for different values of the exploration rate $\epsilon$. The analysis results for the first transition point $\hat{r}_{1}^{*}$ from Eq.~(\ref{eq:transition_point1}) are indicated by red arrows, which are consistent with the simulation results. The insets provide a close-up view of the transition points for clearer observation of the analysis results. The rest learning parameters in (a) are 
		$\alpha = 0.1$ and $\epsilon = 0.01$, while in (b) are $\alpha = 0.1$ and $\gamma = 0.9$. The scales of population in (a), (b) are $|\mathcal{N}| = 30\times 30$.
	}
	\label{fig:transition_point_analysis}
\end{figure}

In our PGG setting, we have $\Pi_{01} = \hat{r}-1$, $\Pi_{10} = \hat{r}$ and $\Pi_{00} = 0$. Due to the competitive balance between $Q_{10,0}$ and $Q_{10,1}$ at the transition point $r_{1}^{*}$, we find that  $Q_{10,0}^* = Q_{10,1}^*$ at $r_{1}^{*}$, i.e,  
\begin{eqnarray}\label{eq:transition_point1}
	\hat{r}^{*}_{1} \approx \frac{(1 - \frac{\epsilon}{2})}{(1 - \frac{\epsilon}{2}) + \gamma (1 - \frac{\epsilon}{2})^{|e|}}.
\end{eqnarray}
According to the above equation, one learns that when $\hat{r}<\hat{r}^{*}_{1}$, the self-looped $00\leftrightarrow 00$ mode dominates over the cyclic $01\leftrightarrow 10$ mode, resulting in the absence of cooperation. On the other hand, when $\hat{r}\gtrsim\hat{r}^{*}_{1}$, $01\leftrightarrow 10$ mode takes over $00\leftrightarrow 00$ mode, leading to the emergence of cooperation.
To verify our analysis, we provide more simulated results obtained under various values of $\gamma$ and $\epsilon$ in Fig.~\ref{fig:transition_point_analysis}. The results are consistent with our analysis, which validates the approach.

Equation (\ref{eq:transition_point1}) from the analysis reveals that a population with a high $\gamma$, indicative of long-sightedness, and low exploration is more likely to respond with kindness when other agents occasionally exhibit kindness. This reciprocal kindness facilitates the achievement of cooperation within the population more readily. Furthermore, the analysis indicates that the phase transition at $\hat{r}_{1}^*$ is continuous, demonstrating that the emergence of cooperation occurs gradually rather than abruptly as $\hat{r}$ increases. Additionally, this conclusion is further supported by the observed finite-size effects of simulation, that is $\bar{f}_c$ increases with the decrease of even-sized $|\mathcal{N}|$ as $\hat{r}\rightarrow \hat{r}^{*+}_{1}$ [see Fig.~\ref{fig:global_fc}].

\subsection{Chessboard Structure}\label{subsec:chessbord}
\subsubsection{State Distribution and Cooperation Preferences across Different States}\label{subsubsec:distribution}
In this section, we will explore the chessboard structure of the spatial pattern based on further simulations. It is important to clarify that this structure is a novel spatial phenomenon that emerges through the accumulation of time [see Fig.~\ref{fig:local_fc} and Fig.~\ref{fig:app_pattern1}]. This is distinct from previous works where such patterns are observed at each step~\cite{ding2024emergence,wang2006memory,shen2024learning}. Therefore, as discussed in Sec.~\ref{subsec:first_transition_point}, it is crucial to focus initially on the distribution of states over time and the averaged cooperation preference for these states over time. However, the emergence of the chessboard structure suggests that agents’ behavior exhibits a spontaneous \emph{regular symmetry breaking} in space, which is distinct from the behavior observed around $\hat{r}^{*}_{1}$ in Sec.~{\ref{subsec:first_transition_point}}. 

The breaking of spatial symmetry indicates that the population contains two categories based on the average cooperation levels, $\mathcal{N}_{h}$, where the cooperation level within their hyperedges is higher than in their neighbors’ hyperedges, and $\mathcal{N}_{l}$, where it is lower. The definitions for $\mathcal{N}_{h}$ and $\mathcal{N}_{l}$ are 
\begin{subequations} \label{eq:N_hl}
	\begin{align}
 	\mathcal{N}_{h} :=  \sum\limits_{{i}\in \mathcal{N}}\mathbbm{1}_{\bar{f}_{c}^{i} > \bar{f}_{c}^{j}, \forall e^j\in \{e|i\in e\}_{-e^{i}}}, \\
 	\mathcal{N}_{l} :=  \sum\limits_{{i}\in \mathcal{N}}\mathbbm{1}_{\bar{f}_{c}^{i} < \bar{f}_{c}^{j}, \forall e^j\in \{e|i\in e\}_{-e^{i}}}.
	\end{align}
\end{subequations}  
Here, $\{e|i\in e\}_{-e^{i}}$ denotes the set of neighboring hyperedges of $e^{i}$. 
Then, we can further define the probability for a given state $s$ in $\mathcal{N}_{h}$ ($\mathcal{N}_{l}$) is
\begin{eqnarray}\label{eq:fs2}
	\bar{f}_{s}^{h(l)} := \frac{\sum\limits_{\tau = t_0}^{t}\sum\limits_{i\in\mathcal{N}_{h(l)}}\mathbbm{1}_{s^{i,i}(\tau) = s}}{|\mathcal{N}_{h(l)}|(t-t_{0})}, 
\end{eqnarray}
and the average cooperation preference for a given state $s$ is 
\begin{eqnarray}\label{eq:phi2}
	\bar{\phi}^{h(l)}_{c|s} := \frac{\sum\limits_{\tau = t_0}^{t}\sum\limits_{i\in\mathcal{N}_{h(l)}} \mathbbm{1}_{s^{i,i}(\tau) = s,a^{i,i}(\tau) = 1}}{\sum\limits_{\tau = t_0}^{t}\sum\limits_{i\in\mathcal{N}_{h(l)}}\mathbbm{1}_{s^{i,i}(\tau) = s}}.
\end{eqnarray}    
Inspired by the inference from Sec.~\ref{sec:simulation}, we select game parameters based on the effect of the local chessboard structure on global cooperation. Specifically, in Case \RNum{1} and Case \RNum{2}, $\hat{r} = 0.72$ is chosen from the range $\hat{r}^*_1<\hat{r}<\hat{r}^*_2$ where $\mathcal{N}_{h}^{\text{\RNum{1}}}\cup \mathcal{N}^{\text{\RNum{1}}}_{l} = \mathcal{N}$, and $\hat{r} = 0.84$ is chosen from the range $\hat{r}>\hat{r}^*_2$ where $\mathcal{N}^{\text{\RNum{2}}}_{h}\cup \mathcal{N}^{\text{\RNum{2}}}_{l} \subsetneq \mathcal{N}$.

\begin{figure}[htbp!]
	\centering
	\includegraphics[width=0.9\linewidth]{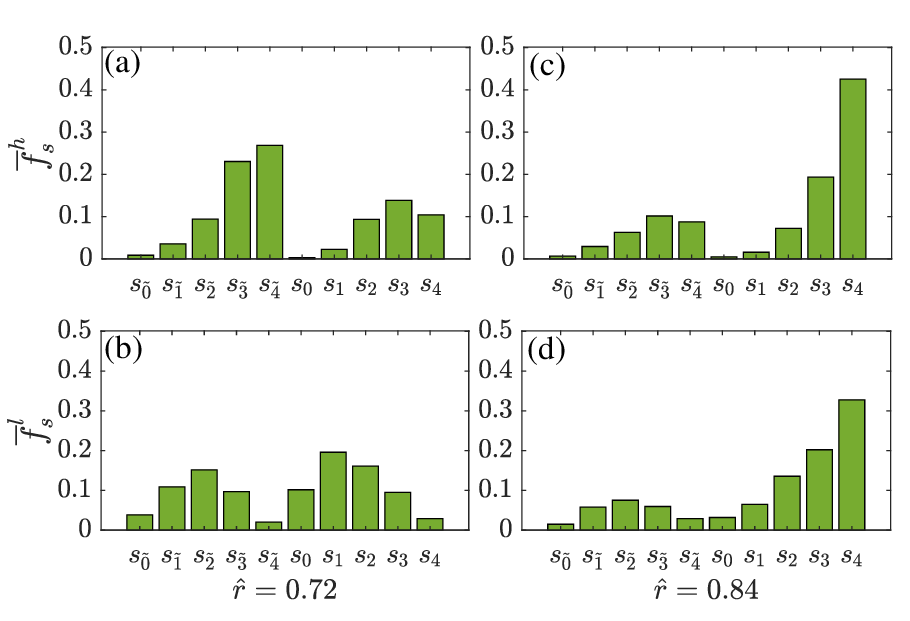}	
	\caption{ (Color online) {\bf The distribution of states for initiators within different categories.} Panels (a, b) and (c, d) show $\bar{f}_{s}^{h}$ and $\bar{f}_{s}^{l}$ for any $s\in\mathcal{S}$ within the categories $\mathcal{N}_{h}$ and $\mathcal{N}_{l}$ under the selected game parameters in Cases \RNum{1} and \RNum{2}. (a, b) display there is a significant difference between the state distribution within $\mathcal{N}_{h}$ and $\mathcal{N}_{l}$ in Case \RNum{1}, while such difference gradually decreases in Case \RNum{2} as shown in (c, d). In (a - d), the population size $|\mathcal{N}| = 30\times 30$, and other parameters are the same as Fig.~\ref{fig:local_fc}.
	}
	\label{fig:fs}
\end{figure}
In Fig.~\ref{fig:fs} (a - d), we present $\bar{f}_{s}^{h(l)}$ of Eq.~\eqref{eq:fs2} within  $\mathcal{N}_h$ and $\mathcal{N}_l$ in the above cases. In the Case \RNum{1}, as shown in (a) and (b) for $\mathcal{N}^{\text{\RNum{1}}}_{h}$, the probability of initiators acting as contributors, $\sum_{s_m} f^{h}_{s_m} = 0.362$, is lower compared to their probability as free-riders, $\sum_{s_{\tilde{m}}} f^{h}_{s_{\tilde{m}}} = 0.638$. In contrast, in  $\mathcal{N}^{\text{\RNum{1}}}_{l}$, the probability of initiators acting as contributors, $\sum_{s_m} f^{l}_{s_m} = 0.5835$, is higher than their probability of acting as free-riders, $\sum_{s_{\tilde{m}}} f^{l}_{s_{\tilde{m}}} = 0.4165$. Additionally, this observation, $\sum_{s_m} f^{h}_{s_m} < \sum_{s_{m}} f^{l}_{s_m}$, aligns with the conclusion drawn from Fig.~\ref{fig:local_fc} (f). That is the initiators’ preference for cooperation is negatively correlated with the local cooperation level within their hyperedges.
While in Case \RNum{2}, (c) and (d) display the negative correlation still holds, i.e., $\sum_{s_m} f^{h}_{s_m} \lesssim \sum_{s_{m}} f^{l}_{s_m}$. However, in $\mathcal{N}^{\text{\RNum{2}}}_{h}$, the probability of initiators acting as contributors becomes higher than that of those acting as free-riders, i.e., $\sum_{s_m} f^{h}_{s_m} > \sum_{s_{\tilde{m}}} f^{h}_{s_{\tilde{m}}}$. This is a difference between Case \RNum{1} and Case \RNum{2} in the state distribution.

Additionally, for Case \RNum{1} in both $\mathcal{N}^{\text{\RNum{1}}}_{h}$ and $\mathcal{N}^{\text{\RNum{1}}}_{l}$, the probability of observing a given state for the contributing initiators varies non-monotonically with the number of contributors in that state. Specifically, both $f^{h}_{s_{m}}$ and $f^{l}_{s_{m}}$ first increases and then slightly decreases with the increase of $m$. However, for a state of a free-riding initiator, the non-monotonicity of $f_{s_{\tilde m}}^{l}$ with $m$ persists, while $f_{s_{\tilde m}}^{h}$ changes to a monotonically increase with $m$ [see Fig.~\ref{fig:fs} (a, b)]. The results suggest the spatial symmetry breaking also emerges in state distribution for Case \RNum{1}, especially for the free-riding initiators. However, in Case \RNum{2}, $f_{s_{\tilde m}}^{h}$ and $f_{s_{\tilde m}}^{l}$ transition to non-monotonic behavior with respect to $m$, whereas $f_{s_{m}}^{h}$ and $f_{s_{m}}^{l}$ increase monotonically with $m$ [see (c, d)]. Thus, it can be concluded that the spatial symmetry breaking observed in the state distribution for Case \RNum{1} weakens significantly in Case \RNum{2}.     

\begin{figure}[htbp]
	\centering
	\includegraphics[width=0.9\linewidth]{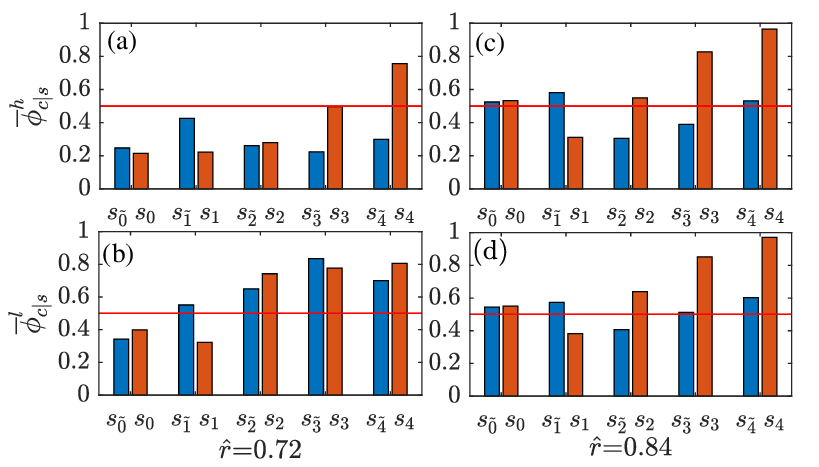}	
	\caption{ (Color online) {\bf Cooperation preferences of initiators across various states within different categories.} (a, b) and (c, d) respectively show the cooperation preferences for initiators within $\mathcal{N}_{h(l)}$, as denoted by $\bar{\phi}_{c|s}^{h(l)}$ in Eq.~(\ref{eq:phi2}), in Cases \RNum{1} and \RNum{2}. In (a - d), the scale of population size $|\mathcal{N}| = 30\times 30$, and the other parameters are consistent with those in Fig.~\ref{fig:local_fc}.   
	}
	\label{fig:phi}
\end{figure}  
Next, Fig.~\ref{fig:phi} (a, b) and (c, d) display $\bar{\phi}^{h(l)}_{c|s}$ defined in Eq.~\eqref{eq:phi2} for any $s\in\mathcal{S}$ under Cases \RNum{1} and \RNum{2}, respectively. In Case \RNum{1}, there is a noticeable gap between $\bar{\phi}^{h}_{c|s}$ and $\bar{\phi}^{l}_{c|s}$ for any $s\in\mathcal{S}$ [see (a, b)]. This gap indicates that initiators in $\mathcal{N}^{\text{\RNum{1}}}_{h}$ have a lower preference for cooperation compared to those in $\mathcal{N}^{\text{\RNum{1}}}_{l}$. In other words, within a hyperedge, the cooperation preference of each initiator is inversely related to that of the participants, exhibiting anti-coordinated. In contrast, in Case \RNum{2}, the difference between $\bar{\phi}^{h}_{c|s}$ and $\bar{\phi}^{l}_{c|s}$ almost entirely vanishes [see (c, d)]. This observation suggests that the spatial symmetry breaking shown in Fig.~\ref{fig:fs} (a, b) is due to differences in cooperation preferences among initiators in $\mathcal{N}^{\text{\RNum{1}}}_{h}$ and $\mathcal{N}^{\text{\RNum{1}}}_{l}$. Furthermore, panels (a) and (c, d) reveal a noticeable gap between $\bar{\phi}^{h(l)}_{c|s_{m}}$ and $\bar{\phi}^{h(l)}_{c|s_{\tilde{m}}}$, whereas this gap is significantly reduced in panel (b). This result suggests that the reliance on previous personal actions in decision-making plays a crucial role in maintaining a high level of cooperation.


\subsubsection{Cohen’s Kappa Coefficient Matrix for Temporal Correlations}\label{subsubsec:kappa2}
Finally, to determine the local state transition modes in the hyperedges, we also investigate Cohen’s kappa coefficient matrix $[\kappa^{h(l)}(s, s^{\prime})]$ in $\mathcal{N}_{h}$ and $\mathcal{N}_{l}$. In this matrix, the element $\kappa^{h(l)}(s, s^{\prime})$ is Cohen’s kappa coefficient for the temporal correlations between consecutive states $s$ and $s^{\prime}$ in $\mathcal{N}_{h}$ and $\mathcal{N}_{l}$, as defined in the following
\begin{eqnarray}\label{eq:kappa2}
	\kappa^{h(l)}(s,s^{\prime}):= \frac{\bar{f}^{h(l)}(s,s^{\prime})-\bar{f}^{h(l)}(s)\bar{f}^{l(h)}(s^{\prime})}{1-\bar{f}^{h(l)}(s)\bar{f}^{h(l)}(s^{\prime})}.
\end{eqnarray}
Here,
\begin{eqnarray}
	\bar{f}^{h(l)}(s,s^{\prime}):=  \frac{\sum\limits_{t=t_0}^{t}\sum\limits_{i\in\mathcal{N}_{h(l)}}\mathbbm{1}_{s^{i,i}(\tau)=s,s^{i,i}(\tau+1)=s^{\prime}}}{|\mathcal{N}_{h(l)}|(t-t_0)}. \nonumber
\end{eqnarray}

\begin{figure}[htbp]
	\centering
	\includegraphics[width=0.9\linewidth]{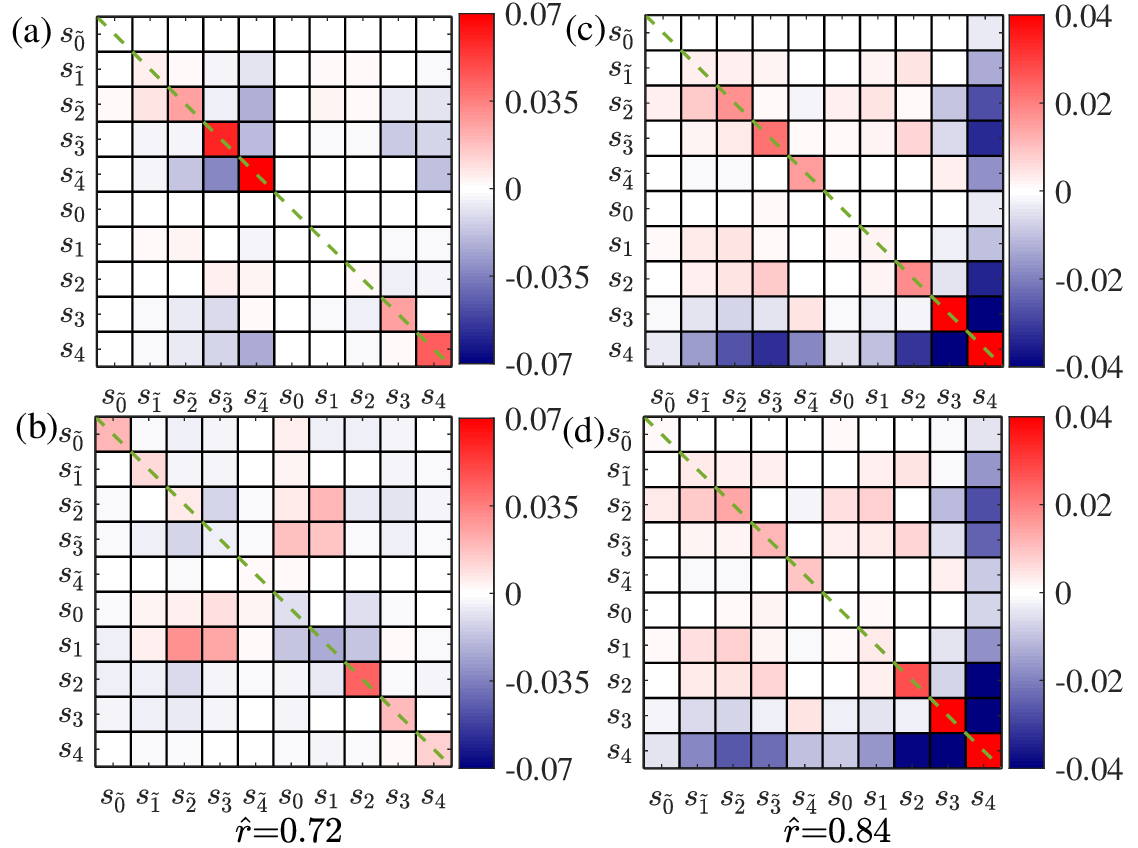}
	\caption{{\bf Cohen’s kappa coefficient matrices for temporal correlations within different categories.} 
		Panels (a, b) show kappa coefficients matrix $[\kappa^{h(l)}(s, s^{\prime})]$ for temporal correlations between consecutive states within $\mathcal{N}_{h}$ and $\mathcal{N}_{l}$, as denoted by $\kappa^{h}(s,s^{\prime})$ and $\kappa^{l}(s,s^{\prime})$ in Case \RNum{1}. Panels (c, d) illustrate the corresponding coefficient matrix in Case \RNum{2}.
		In (a - d), the averages are computed over $10^8$ steps, after $10^{10}$ transient steps.
	}\label{fig:kappa2}
\end{figure}
In Fig.~\ref{fig:kappa2}, we show $[\kappa^{l(h)}(s, s^{\prime})]$ in Case \RNum{1} and Case \RNum{2}. The difference of $[\kappa^{l(h)}(s, s^{\prime})]$ in between (a) and (b) indicates that spatial symmetry breaking is also reflected in the temporal correlation between consecutive states in Case \RNum{1}. In $\mathcal{N}^{\text{\RNum{1}}}_{h}$, self-loop transitions prevail in the hyperedges, with $s_{\tilde 4}\leftrightarrow s_{\tilde 4}$ and $s_{\tilde{3}}\leftrightarrow s_{\tilde 3}$ modes being the most dominant [see (a) and Fig.~\ref{fig:app_temporal_mode} (a2) ].
In these dominant modes, the initiators free-ride on the contributions of participants. 
In contrast, in $\mathcal{N}^{\text{\RNum{1}}}_{l}$, these free-riding modes disappear due to the emergence of various competing cyclic modes but the self-looped synergy modes, such as $s_{4}\leftrightarrow s_{4}$ and $s_{3}\leftrightarrow s_{3}$, still persist [see (b) and Fig.~\ref{fig:app_temporal_mode} (b1 - b3)]. Additionally, a new anti-coordinated and self-looped mode emerges, specifically $s_{2}\leftrightarrow s_{2}$.

In Case \RNum{2}, Fig.~\ref{fig:kappa2} (c, d) shows that the coefficient matrices for $\mathcal{N}^{\text{\RNum{2}}}_{h}$ and $\mathcal{N}^{\text{\RNum{2}}}_{l}$ retain part of the features of the coefficient matrices for $\mathcal{N}^{\text{\RNum{1}}}_{h}$ and $\mathcal{N}^{\text{\RNum{1}}}_{l}$, respectively. However, 
the spatial symmetry breaking observed in the coefficient matrix for Case \RNum{1} is also significantly reduced. Furthermore, both in $\mathcal{N}^{\text{\RNum{2}}}_{h}$ and $\mathcal{N}^{\text{\RNum{2}}}_{l}$, the self-loop modes are still dominant, similar to $\mathcal{N}^{\text{\RNum{1}}}_{h}$. However, the most dominant self-loop modes are synergistic cooperation modes rather than free-riding modes in $\mathcal{N}^{\text{\RNum{1}}}_{h}$ [see (a), (c, d) and Fig.~\ref{fig:app_temporal_mode} (c1)]. In addition, similar to those in $\mathcal{N}^{\text{\RNum{1}}}_{l}$, a significant number of cyclic modes continue to persist in both $\mathcal{N}^{\text{\RNum{2}}}_{h}$ and $\mathcal{N}^{\text{\RNum{2}}}_{l}$ [see (b), (c, d) and Fig.~\ref{fig:app_temporal_mode} (c2)]. 
Based on these analyses, a plausible conjecture arises: the cyclic modes may contribute to the destabilization of the free-riding modes, potentially leading the dominant modes to transition into the synergy modes. Notably, this transformation appears to enhance the level of global cooperation. However, it is crucial to recognize that this shift may also diminish the local cooperation of the cyclic modes within their hyperedges as a sacrifice.

\section{Discussion and Conclusion}\label{sec:summary} 
In this study, we propose a new model of the other-regarding reinforcement learning evolutionary game and apply it to the PGG on von Neumann hyperedges. Interestingly, unlike only one transition observed in traditional evolutionary games or the self-regarding reinforcement learning case~\cite{wang2023synergistic,shen2024learning}, the average cooperation level in our model experiences two sharp increases as the synergy factor is varied. The interval of the synergy factor is divided into three distinct regions (\emph{absence of cooperation}, \emph{medium cooperation}, and \emph{high cooperation}) connected by two transition points. 
Furthermore, based on the spatial patterns observed in these regions, we identify the regular and anti-coordinated chessboard structures in the spatial pattern that positively contribute to the first sharp increase in transition but adversely affect the second one. Additionally, this structure develops through the accumulation of time rather than emerging at each step as in the previous works ~\cite{shen2024learning,ding2024emergence,wang2006memory}. 

We further reveal that there is a competing balance between two state transition modes at the first transition point: whether a free-rider will reciprocate the kindness of another contributor or continue to free-ride in response to such kindness. Following this clue, we give the theoretical transition point for the emergence of cooperation, and reveal a population with a long-sighted perspective and low exploration rate is more likely to reciprocate kindness with each other,  ultimately promoting cooperation. 

Finally, for the chessboard structures, we investigate the state distribution and transition modes for initiators within different categories classified by local symmetry breaking. 
The state distribution illustrates that agents’ attention to their own historical information is essential for promoting cooperation when making decisions to act. The transition modes indicate the free-riding modes, in which the initiators free-ride on the contributions of participants, play a significant role in the stability of the local chessboard structure. However, as the synergy factor increases, these modes are gradually replaced by the synergy modes due to the erosion of competing cyclic modes. 

Although our model reveals new mechanisms for the emergence of cooperation in the context of PGGs, several open questions remain. One issue is whether the local regular symmetry breaking and anti-coordination phenomena persist in more complex hypergraphs. Furthermore, our theory provides the conditions for the emergence of cooperation based on competition involving only two modes. However, developing a comprehensive theory that considers competition involving multiple modes, such as the second transition point, or heterogeneous hypergraphs remains a challenge. Additionally, similar to our previous work~\cite{ding2024emergence,zheng2024evolution}, the computational complexity of RLEGs complicates the identification of phase transitions and transition points ~\cite{Zhao2024Emergence,domb2000phase,stanley1971phase}, particularly the finite-size scaling~\cite{privman1990finite,brezin1982investigation}. Therefore, designing more efficient algorithms remains a crucial challenge for simulating large-scale simulations. In conclusion, addressing these questions could guide further research efforts and enhance our understanding of collective behaviors, particularly regarding the evolution of grouping cooperation within the framework of reinforcement learning.

\section*{ACKNOWLEDGEMENTS}
We thank Weiran Cai, Shengfeng Deng, and Bin-Quan Li for their constructive suggestions on this research. We are supported by the Natural Science Foundation of China under Grants No. 12165014 and 12075144, and the Key Research and Development Program of Ningxia Province in China under Grant No. 2021BEB04032.

\appendix
\renewcommand{\thefigure}{\Alph{section}.\arabic{figure}} 
\section{The Model for Evolutionary Game}\label{sec:app_EG_model}
Here, we introduce the traditional evolutionary games (EGs) in the context of PGGs, applied to hypergraph generated from von Neumann lattices [see Sec.~\ref{sec:simulation}]. Similar to OR-RLEGs, $|\mathcal{N}|$ agents are also positioned on an $L \times L$ hypergraph, and each evolutionary step includes gaming and learning processes.
In gaming process, any agent $i$ will initiate a public goods game in a hyperedge $e^{i}$ and the other agents within $e^{i}$ will participate in it. However, unlike OR-RLEGs, in this model, each agent takes the same action $a^{i}$ across all the games it is involved in at each step. This action $a^i$ is either its own action $\tilde{a}^{i}$, chosen with probability $1-\mu$ or a random action from $\mathcal{A}$, chosen with mutation probability $\mu$. After the action-takings, the initiator $i$ within hyperedge $e^{i}$ receives its payoff at $\tau$, which is determined based on the actions taken by the agents within $e^i$ as follows
\begin{eqnarray}\label{eq:app_payoff}
	\Pi^{i}(\tau) = \frac{r}{|e^i|}\cdot\sum\limits_{j \in e^i} a^{j}(\tau) - a^{i}(\tau).
\end{eqnarray}

During the learning process, each initiator $i$ randomly selects another initiator $j$ in the neighboring hyperedge, as the model agent to update its own action according to Fermi rule~\cite{tanimoto2015fundamentals}. The details are as follows:
\begin{eqnarray}\label{eq:app_fermi_rule}
	\tilde{a}^{i}(\tau+1) = 
	\left\{
	\begin{array}{lll}
		a^j(\tau) &\quad& \displaystyle{p = \frac{1}{1+\exp[\beta(\Pi^{i}-\Pi^{j})]} }\\
		\tilde{a}^i(\tau) &\quad& \text{otherwise},
	\end{array}
	\right.
\end{eqnarray}
which denotes that the agent $i$ either replaces its own action $\tilde{a}^{i}$ with $j$'s action $a^{j}$ at current round or retains its original action $\tilde{a}^i$. Here, to maintain consistency with OR-RLEGs, we set the agents to focus only on the reward when they act as initiators rather than as participants in the process of optimizing cognition. Furthermore, simulation takes a synchronous update rather than the traditional asynchronous update approach~\cite{hauert2010replicator}.

To summarize, the pseudo-code is provided in Algorithm~\ref{algorithm:protocol2}.

\begin{algorithm}[htbp!]  
	\caption{The Algorithm of Evolutionary Public Goods Game}\label{algorithm:protocol2}
	\LinesNumbered 
	\KwIn{Learning parameters: $\beta$, $\mu$; Population:$\mathcal{N}$}
	{\bf Initialization}\
	\For{$i$ in $\mathcal{N}$}{ 
		Pick an action $a^i$ randomly from $\mathcal{A}$ for public games \	
	}
	\Repeat{\text the system becomes statistically stable or evolves for the desired time duration}
	{{\bf Gaming process}\;
		\For{$i$ in $\mathcal{N}$}{ 
			Initiate a public good game consist of all agents in $e^i$ \
			\For{$j$ in $e^i$}{
				\KwResult{ActionTaking}
				$ran \gets \text{RandomNumber()}$ \; 
				\eIf{$ran < \mu$}{
					Takes a random mutation action from $\mathcal{A}$\;
				}{
					Takes its own action $\tilde{a}^j$\;
				}
			}				
			Get reward $\Pi^i$ according to Eq.~(\ref{eq:app_payoff})\;	
		}
		{\bf Learning process}\;
		\For{$i$ in $\mathcal{N}$}{
			$j\gets \text{RandomInitiator in the neighboring hyperedge} $ \;
			Update action $\tilde{a}^i$ according to Eq.~(\ref{eq:app_fermi_rule})			
		}
	}
\end{algorithm}

\section{Further Simulation for Local Cooperation Levels}
\subsection{The Spatial Pattern}\label{subsec:app_pattern}
\begin{figure}[htbp!]
	\centering
	\includegraphics[width=\linewidth]{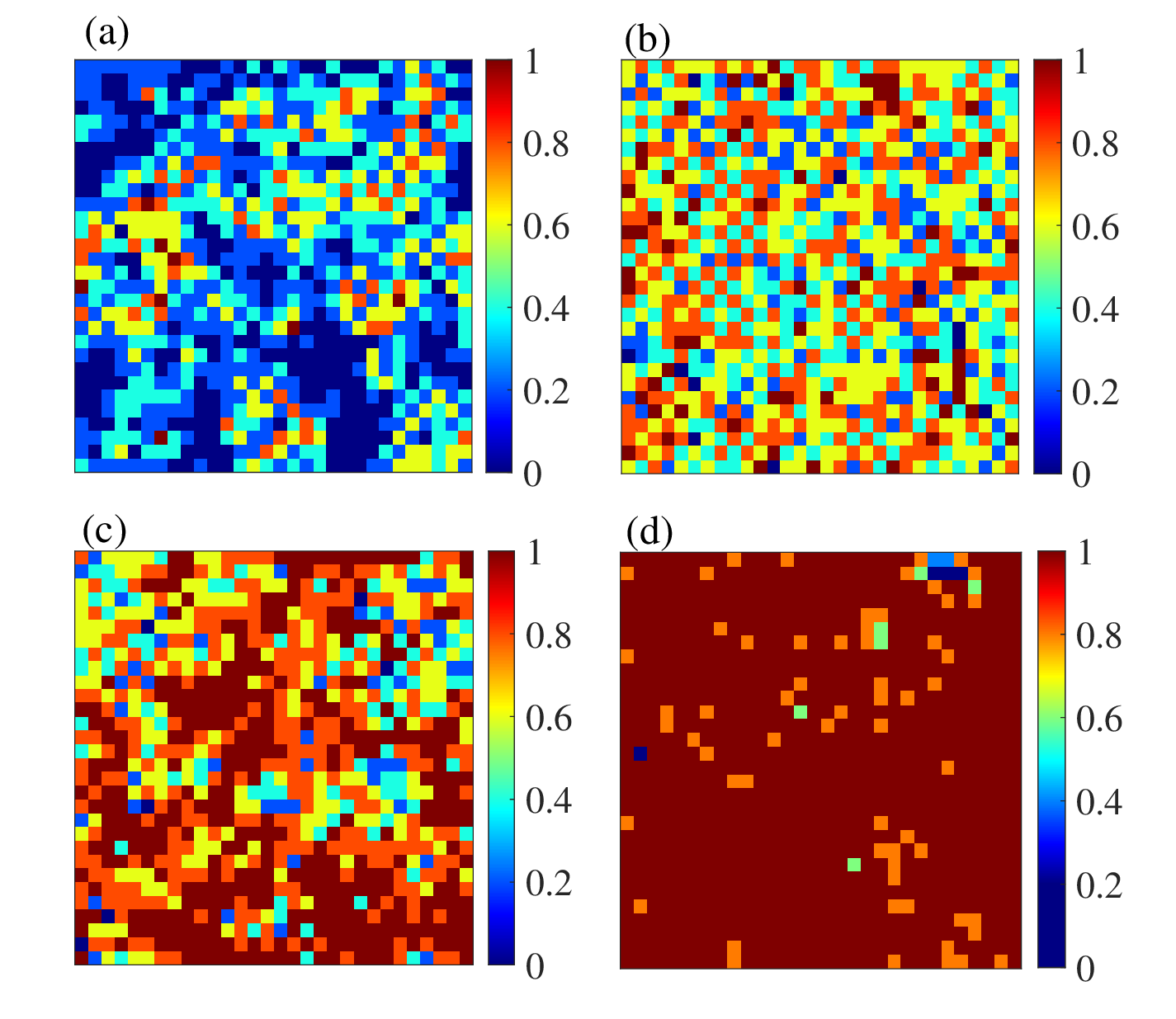}
	\caption{{\bf Spatial patterns of local cooperation at a specific step.} In the figure, we present the spatial pattern of $f_{c}^{i}$ for each hyperedge at $\tau = 10^{9}$ on the von Neumann hypergraph. (a - d) display the game parameters $\hat{r} = 0.54, 0.72, 0.84$, and $1.00$, respectively, which correspond to those in (b - e) of Fig.~\ref{fig:local_fc}. The population size is $|\mathcal{N}|=30\times 30$.
	}\label{fig:app_pattern1}
\end{figure}

Here, we firstly explore local cooperation level on the von Neumann hypergraph from spatial patterns in more detail. Fig.~\ref{fig:app_pattern1} shows the pattern of $f_{c}^{i}$ for each hyperedge $e^{i}\in\mathcal{E}$ at a specific step within the stable duration. Compared to the spatial pattern of the average local cooperation level over time shown in Fig.~\ref{fig:local_fc}, the chessboard structure becomes less pronounced for the local cooperation level at a specific step. Especially in the case $\hat{r} = 0.72$, the chessboard structure is both regular and global in the spatial pattern for $\bar{f}_{c}^{i}$ [see Fig.~\ref{fig:local_fc}(c) in Sec.~\ref{sec:simulation}]. However, the feature of this structure completely disappears for 
$f_{c}^{i}$ at this specific step [see Fig.~\ref{fig:app_pattern1} (b)]. In other words, the chessboard pattern is a spatial phenomenon that emerges through time accumulation rather than appearing at each step. 

\begin{figure}[htbp]
	\centering
	\includegraphics[width=\linewidth]{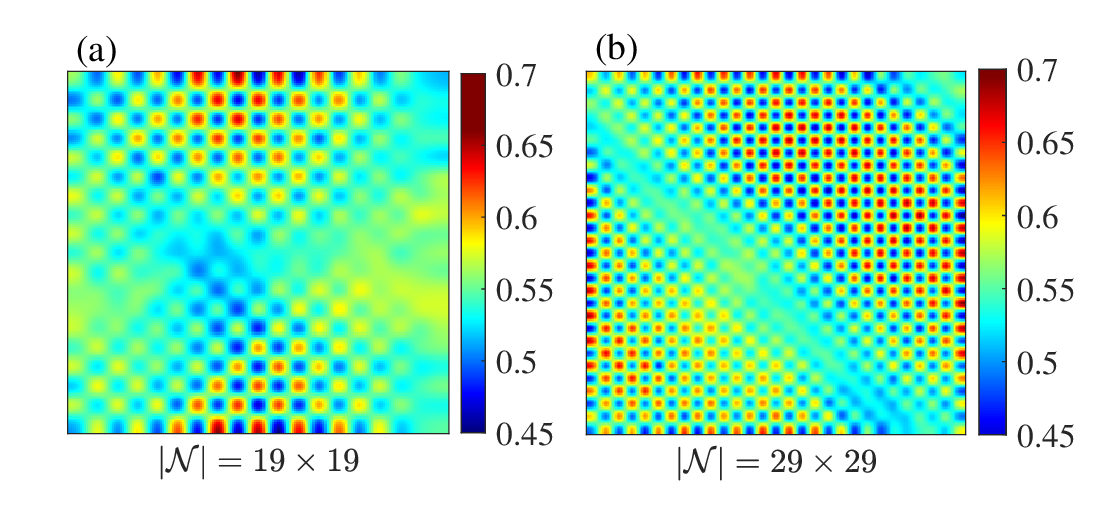}
	\caption{{\bf Spatial patterns of averaged cooperation level locally in different odd-sized von Neumann hypergraphs.} The figure displays the spatial patterns of the average local cooperation preference $\bar{f}_{c}^{i}$ in different odd-sized von Neumann hypergraphs. The scales of hypergraphs in (a) and (b) are $|\mathcal{N}|=19\times 19$ and $29\times 29$, respectively. The average are computed over $10^8$ steps, after $10^{10}$ transient steps.
	}\label{fig:app_pattern2}
\end{figure}

In Fig.~\ref{fig:app_pattern2}, we present the spatial pattern of the average local cooperation level on different odd-sized von Neumann hypergraphs. The figure illustrates that fragmented regions of the chessboard structure appear along the diagonal of the network, and the proportion of these regions increases with the decrease of $|\mathcal{N}|$. This suggests that smaller odd-sized $|\mathcal{N}|$ are more detrimental to the formation of a global chessboard pattern. Consequently, the size of $|\mathcal{N}|$ affects the global cooperation level $\bar{f}_c$ in the MC regions or in the area between MC and HC regions when $|\mathcal{N}|$ is odd-sized [see Fig.~\ref{fig:global_fc}].

\subsection{Stability of Regular Symmetry Breaking in Local}\label{subsec:app_stability}
Given the spatial pattern that emerges over the accumulation of time, we further proceed to examine the time series of cooperation levels within distinct hyperedges. Our focus is to discern whether there are notable temporal patterns that occur within these hyperedges. To uncover these patterns, we define the cooperation levels within any hyperedge $e^{i}$ over a sliding time window with a scale of $\Delta t$ as
\begin{eqnarray} \label{eq:hat_fc}
	\hat{f}_{c}^{i}(\tau) := \frac{\sum\limits_{\tau^{\prime}=\tau-\Delta t}^{\tau}f_{c}^{i}(\tau^{\prime})}{\Delta t},
\end{eqnarray}
and investigate $\hat{f}_{c}^{i}$ over the time series.
According to the definitions provided in Sec.~\ref{subsec:chessbord}, our game parameters are set as $\hat{r} = 0.6$ and $0.72$, falling within Case \RNum{1} with $\hat{r}^*_1<\hat{r}<\hat{r}^*_2$. Additionally, we set $\hat{r} = 0.84$, falling within Case \RNum{2} with $\hat{r}>\hat{r}^*_2$. Furthermore, because of the spatial symmetry-breaking in pattern, we display time series of $\hat{f}^{i}_{c}$ for a pair of neighboring hyperedges, where one has a low and the other has a high $\bar{f}_{c}^{i}$ over the entire time duration. 
\begin{figure}[htbp]
	\centering
	\includegraphics[width=\linewidth]{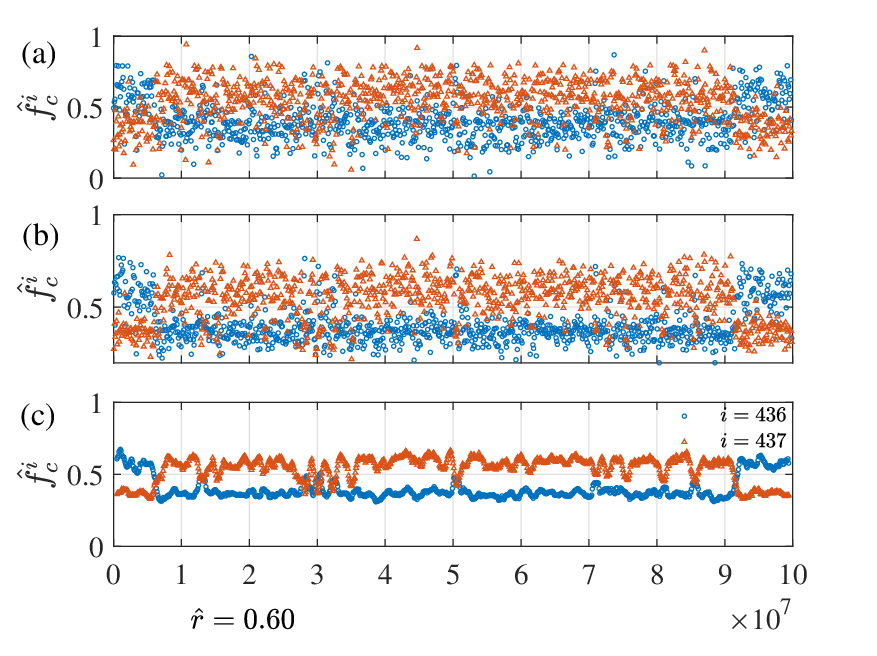}
	\caption{{\bf Time series of the local cooperation levels across various sliding window scales.} In the figure, (a - c) present the time series of $\bar{f}_{c}^{i}$ within two neighboring hyperedges across various sliding window scales, $\Delta t = 10^4$, $10^5$ and $10^6$. To manage the large number of data points, one data point is displayed for every thousand. In each panel, the red dots indicate a high $\bar{f}_{c}^{i}$, while the blue dots indicate a low $\bar{f}_{c}^{i}$ within the corresponding hyperedge over the entire period. The Pearson correlation coefficients of time series between the neighboring hyperedges in (a - c) are $-0.136$, $-0.421$, and  $-0.838$, respectively. The population scale is $|\mathcal{N}|=30\times 30$.
	}\label{fig:app_timeseries1}
\end{figure}

According to Fig.~\ref{fig:app_timeseries1} (a - b) with $\hat{r} = 0.6$, it is observed that the time series of $\hat{f}_{c}^{i}$ exhibits disorder for small-scale time windows. Whereas for larger time windows in (c),  $\hat{f}^{i}_{c}(\tau)$ exhibits a noticeable gap between the neighboring hyperedges within each window. 
Furthermore, (a - c) shows a multiscale negative correlation for $\hat{f}_{c}^{i}$ between neighboring hyperedges across time windows, demonstrating that the negative correlation increases with time window scale. As (c) shows, this correlation at larger scales may reverse the gap in $\hat{f}^{i}_{c}(\tau)$ between neighboring hyperedges, potentially destabilizing local symmetry breaking. Thus, the chessboard becomes blurry in spatial pattern and clusters of checkerboard structures will emerge [see Fig.~\ref{fig:local_fc}].


\begin{figure}[htbp]
	\centering
	\includegraphics[width=\linewidth]{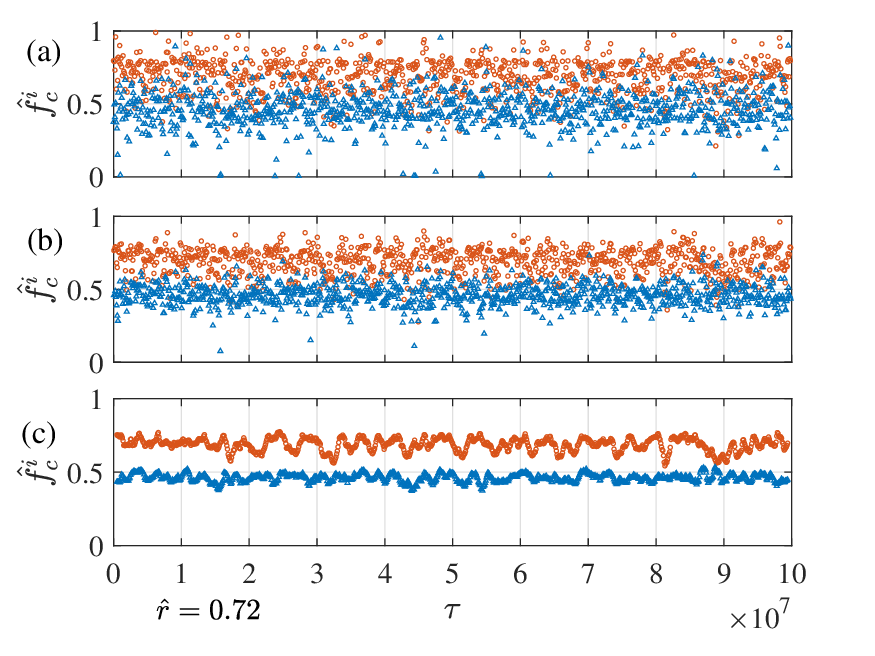}
	\caption{{\bf Time series of the local cooperation level across various sliding window scales.} Panels (a - c) present the time series of $\bar{f}_{c}^{i}$ within two neighboring hyperedges across various sliding window scales. The Pearson correlation coefficients of time series between the neighboring hyperedges in (a - c) are $0.29$, $0.312$, and $0.253$, respectively.
	The settings and display methods in panels (a - c) are consistent with those in Fig.~\ref{fig:app_timeseries1}.
	}\label{fig:app_timeseries2}
\end{figure}

\begin{figure}[htbp]
	\centering
	\includegraphics[width=\linewidth]{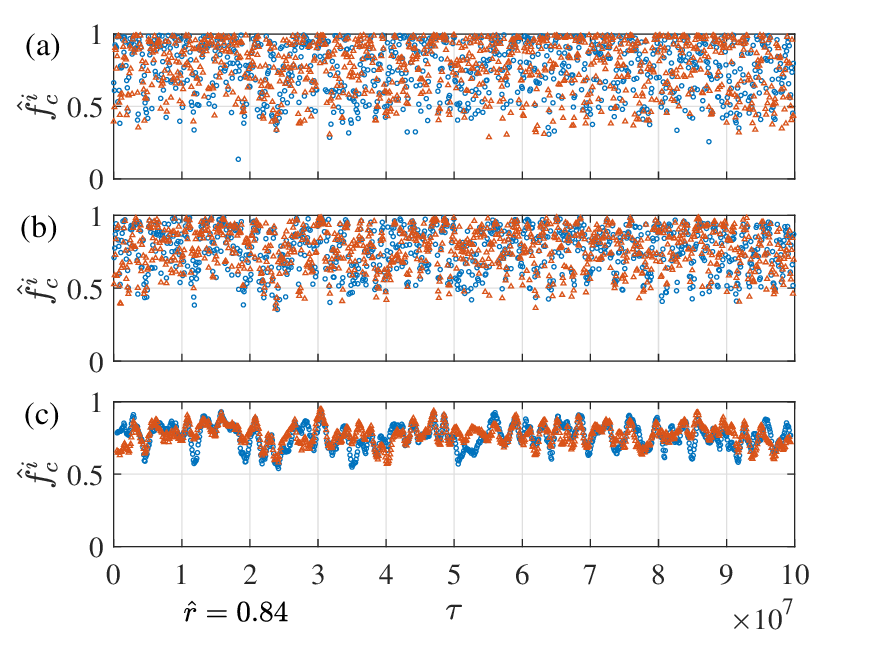}
	\caption{{\bf Time series of cooperation levels across various sliding window scales.} Panels (a - c) present the time series of $\bar{f}_{c}^{i}$ within two neighboring hyperedges across various sliding window scales. The Pearson correlation coefficients of the time series between the neighboring hyperedges in (a - c) are $0.631$,  $0.705$, and $0.569$, respectively. 
	The settings and display methods in panels (a - c) are consistent with those in Fig.~\ref{fig:app_timeseries1}.
	}\label{fig:app_timeseries3}
\end{figure}

With the increase of $\hat{r}$ but still in Case \RNum{1}, Fig.~\ref{fig:app_timeseries2} (a - c) show $\hat{f}_{c}^{i}(\tau)$ retains most features observed in Fig.~\ref{fig:app_timeseries1} for both small-scale and larger time windows. However, different from Fig.~\ref{fig:app_timeseries1}, the previously noted negative correlation is transformed into a positive correlation, which stabilizes local symmetry breaking and extends its characteristic time. This stability allows the chessboard structure to dominate the spatial pattern [see Fig.~\ref{fig:local_fc} (c)]. As $\hat{r}$ continues to increase and transitions into Case \RNum{2}, Fig.~\ref{fig:app_timeseries3} reveals that although the positive correlation
of $\hat{f}^{i}_{c}$ between neighboring hyperedges will further increase, the reduction of the gap still leads to a loss of stability in local symmetry breaking. Consequently, clusters start to re-emerge and the chessboard structure becomes blurry once again [see Fig.~\ref{fig:local_fc} (d)].

In sum, one learns that the stability of local symmetry breaking depends on both the gap in $\bar{f}_{c}^{i}$ between neighboring hyperedges and the correlation of $\hat{f}_{c}^{i}$ between them, especially on a large scale. A small gap and a strong negative correlation both destabilize local symmetry breaking and shorten its characteristic time. This length, in turn, influences the composition and size of clusters consisting of chessboard structures within the spatial patterns. For low $\hat{r}$, the negative correlation plays a main role in destabilizing local symmetry breaking, while for high $\hat{r}$, the narrowing of the gap becomes the primary factor in this destabilization. Then, for the medium $\hat{r}$, the local symmetry breaking is stable, allowing the chessboard structure to dominate the spatial pattern.

\subsection{ Local State Transition Modes}
In the main text, we state that in Cohen’s coefficient matrix $[\kappa(s, s^{\prime})]$, the positive $\kappa(s,s)$ indicates that the transition of $s$ forms a local self-looped $s\leftrightarrow s$ mode. And, the positive $\kappa(s,s^{\prime})$ and $\kappa(s^{\prime},s)$, with $\kappa(s,s^{\prime})\approx \kappa(s^{\prime},s)$, suggest that the transition between $s$ and $s^{\prime}$ forms a local cyclic $s\leftrightarrow s^{\prime}$ mode [see Sec.~\ref{subsubsec:simulation_for_analysis}]. The statement means we can identify the primary local state transition mode by the coefficient matrix. To illustrate the validity of this statement, we present state transition modes for different initiators belonging to distinct categories in Cases \RNum{1} and \RNum{2}, specifically those initiators are in $\mathcal{N}^{\text{\RNum{1}}}_{h(l)}$ and $\mathcal{N}^{\text{\RNum{2}}}_{h(l)}$. The game parameters are set as $\hat{r} = 0.72$ in Case \RNum{1} and $0.84$ in Case \RNum{2} to maintain consistency with Sec.~\ref{subsec:chessbord}. 

\begin{figure*}[htbp!]
	\centering
	\includegraphics[width=\linewidth]{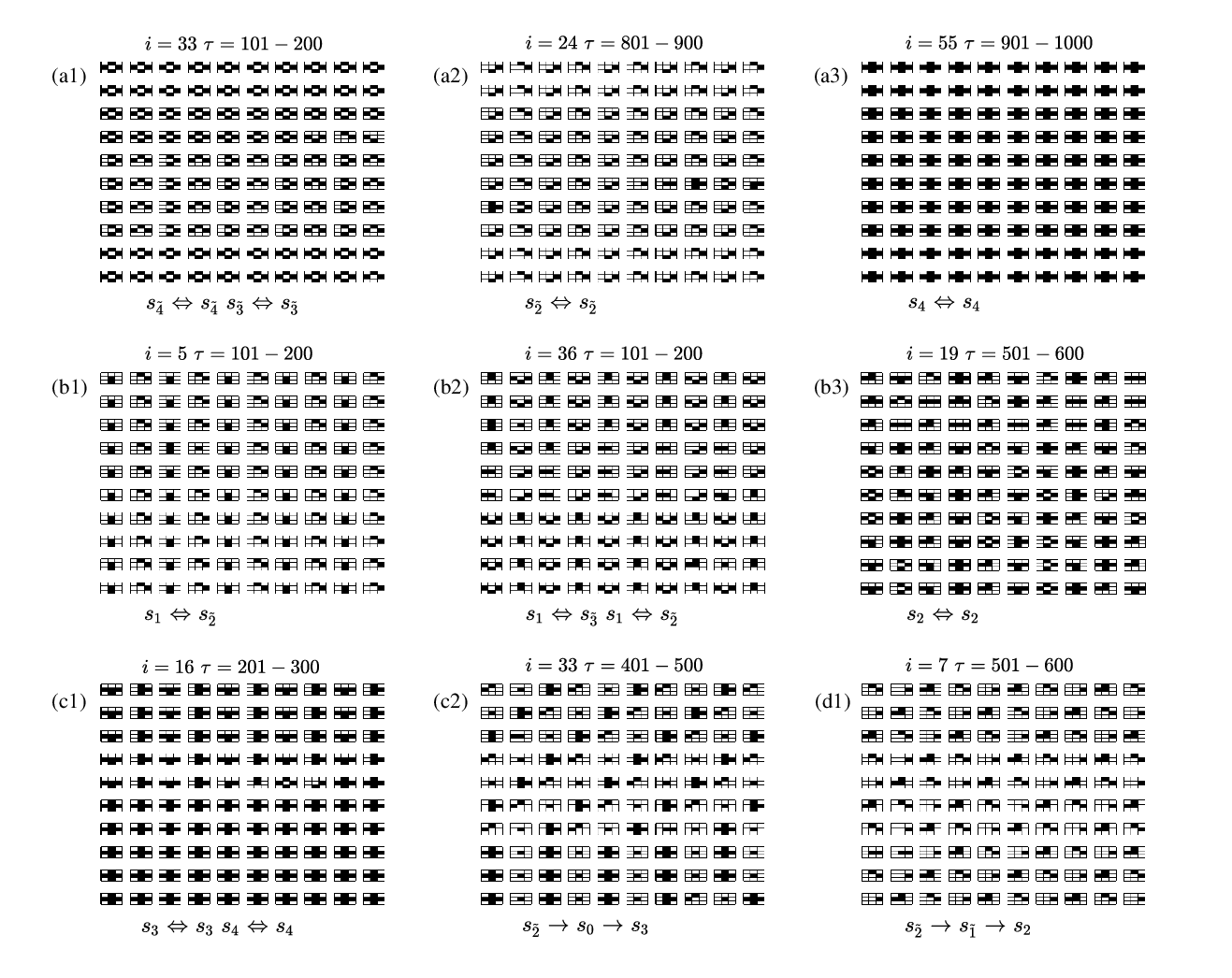}
	\caption{{\bf Time series of local configuration transitions across different specific hyperedges.} In (a1 - a3) and (b1 - b3), we present the time series of local configuration transitions within the hyperedges of initiators in $\mathcal{N}_{h}^{\text{\RNum{1}}}$ and $\mathcal{N}_{l}^{\text{\RNum{1}}}$, respectively. (c1 - c3) and (d1) display the time series within the hyperedge of initiators in $\mathcal{N}_{h}^{\text{\RNum{2}}}$ and $\mathcal{N}_{l}^{\text{\RNum{2}}}$. Here, the hyperedge in (a1 - a3) and (b1 - b3) are selected from the hyperedges of initiators in Fig.~\ref{fig:kappa2} (a) and (b), while that in (c1 - c3) and (d1) from the hyperedges of initiators in Fig.~\ref{fig:kappa2} (c) and (d).
	The main short modes of state transitions for the initiator in each time series are indicated below the corresponding panels. For each configuration in the panels, contributors are filled in black, while free-riders are in white.
	}\label{fig:app_temporal_mode}
\end{figure*}

In Fig.~\ref{fig:app_temporal_mode}, we present the time series of configurations that include all agents’ actions within different hyperedges. The panels in the figure correspond to those in Fig.~\ref{fig:kappa2}, e.g., the parameters and the affiliated categories of initiators in (a1 - a3) are identical to those in panel (a) of Fig.~\ref{fig:kappa2}. In panels (a1 - a3), we can observe the emergence of the main self-loop modes in $\mathcal{N}^{\text{\RNum{1}}}_{h}$ identified by $[\kappa^{l}(s, s^{\prime})]$ of Fig.~\ref{fig:kappa2} (a), such as $s_{\tilde{4}}\leftrightarrow s_{\tilde{4}}$, $s_{\tilde{3}}\leftrightarrow s_{\tilde{3}}$ and $s_{4}\leftrightarrow s_{4}$. 
In addition, based on (a1 - a2), it can be concluded that the period of the configuration transition mode will not be shorter than the period of the state transition mode, as there may be multiple configurations within a single state for the initiators. However, the period of the configuration transition mode can also be reflected in the state transition period of at least one participant in this hyperedge. For instance, the state transition period for both the bottom and top participants is equal to the period of the configuration transition period in (a2). This indicates that the state transition periods for agents within a specific hyperedge may vary. 

In Fig.~\ref{fig:app_temporal_mode} (b1 - b3), the main modes in $\mathcal{N}^{\text{\RNum{1}}}_{l}$ identified by $[\kappa^{l}(s, s^{\prime})]$ in Fig.~\ref{fig:kappa2} (b) are also observed. In $\mathcal{N}^{\text{\RNum{1}}}_{h}$, the dominating self-loop modes in $\mathcal{N}^{\text{\RNum{1}}}_{l}$ disappear and are replaced by shorter cyclic modes, such as $s_{1}\leftrightarrow s_{\tilde{2}}$ and $s_{1}\leftrightarrow s_{\tilde{3}}$. Furthermore, a comparison of panels (b1 - b3) and (a1 - a3) shows that the initiator is more willing to serve as a contributor, while the participants are more likely to adopt the role of free-riders in $\mathcal{N}^{\text{\RNum{1}}}_{l}$. In contrast, the situation is reversed in $\mathcal{N}^{\text{\RNum{1}}}_{h}$. These findings are consistent with the results in Fig.~\ref{fig:phi}.
In Fig.~\ref{fig:app_temporal_mode} (c1), we observe the main modes $s_{3}\leftrightarrow s_{3}$ and
$s_{4}\leftrightarrow s_{4}$ in $\mathcal{N}^{\text{\RNum{2}}}_{l}$, which are identified by Fig.\ref{fig:kappa2} (c). In contrast, in (c2) and (d1), we see some long-period cyclic modes, which are not identified by $[\kappa^{h(l)}(s, s^{\prime})]$. 
Here, we must point out that $[\kappa^{h(l)}(s, s^{\prime})]$ is only able to identify period-2 cyclic modes and self-loop modes. The reason is that long modes are disordered and cannot maintain stability over a long duration, which leads to a weakening of the correlation between consecutive states. However, these long modes are also capable of destroying the stability of some short modes while stabilizing others.
\section{Mathematical Notation Descriptions}\label{sec:notations}
\begin{table*}[htbp!]
	\caption{\label{tab:notation1}%
		{\bf The descriptions for mathematical notations in models}}
	\begin{ruledtabular}
		\begin{tabular}{p{3cm}p{10cm}p{8cm}}
			symbol&  Description & Applied to\\
			\colrule
			$\mathcal{N}$ & The set of all agents in the population &  Both modes\\
			$\mathcal{E}$ & The set consisting of all hyperedges & Both modes\\ 
			$e^{i}$    &   The hyperedge consisting of agent $i$ along with its nearest neighbors in the von Neumann lattice & Both modes\\ 
			$\hat{r}$ & Synergy factor of public goods game & Both modes\\
			$\Pi^{i}(\tau)$  & The payoff for the initiator $i$ at $\tau$ & Both modes\\
			$\mathcal{S}/\mathcal{A}$ & State/action set for agents  &  OR-RLEG\\
			$s^{i,j}/a^{i,j}$  &  The state/action of agent $i$ in the hyperedge $e^{j}$ &  OR-RLEG \\
			$n_{c}^{i,j}$ & The number of contributors other than $i$ in the hyperedge $e^{j}$ &  OR-RLEG \\
			$s_{m}/s_{\tilde{m}}$ & The state for a contributor/free-rider having $m$ other contributors in the corresponding hyperedge &  OR-RLEG \\
			$\alpha$ & Learning rate for agents  &  OR-RLEG\\
			$\gamma $ & Discounting factor for agents &   OR-RLEG\\
			$\epsilon$ & Exploration rate for agents &  OR-RLEG\\
			$\beta$  & Selection intensity &  Traditional EG \\
			$\mu$    & Mutation probability for agents' action &Traditional EG \\
			$\tilde{a}^{i}(\tau)$ & The own action for agent $i$ at $\tau$ & Traditional EG \\
			$a^{i}(\tau)$  & The action that agent $i$ takes at $\tau$ &Traditional EG 
		\end{tabular}
	\end{ruledtabular}
\end{table*}
For easy reference, we provide descriptions of the mathematical notations used in our work, as detailed in Tabs.~\ref{tab:notation1} and \ref{tab:notation2}. Tab.~\ref{tab:notation1} outlines the notations used in the OR-RLEG and traditional EG models, while Tab.~\ref{tab:notation2} presents the notations specific to simulation and analysis.     

\begin{table*}[htbp!]
	\caption{\label{tab:notation2}%
		{\bf The descriptions for mathematical notations in simulation and analysis}}
	\begin{ruledtabular}
		\begin{tabular}{p{3cm}p{10cm}p{4cm}}
			symbol&  Description  & Defined by or in\\
			\colrule
			$f_{c}^{i}(\tau)$ & Local cooperation level in the hyperedge $e^{i}$ at step $\tau$ & Eq.~\eqref{eq:fci}\\
			$f_{c}(\tau)$ & Global cooperation level in the population at step $\tau$ & Eq.~\eqref{eq:fc}\\ 
			$\bar{f}_{c}$    &   Average cooperation level in the global population across the stable stage & Eq.~\eqref{eq:bar_fc}\\ 
			$p^{i}_{c}(\tau)$ & Cooperation preference for agent $i$ at step $\tau$ & Eq.~\eqref{eq:pci} \\
			$\bar{f}_{s}$  &  Probability that initiators in population are in state $s$ & Eq.~\eqref{eq:fs1}\\
			$\bar{\phi}_{c|s}$ & Cooperation preference for initiators in the given state $s$ & Eq.~\eqref{eq:phi1}\\
			$\kappa(s|s^{\prime})$ &  Cohen’s kappa coefficient for the temporal correlation between consecutive states $s$ and $s^{\prime}$ & Eq.~\eqref{eq:kappa1}\\
			$[\kappa(s|s^{\prime})]$ & Cohen’s coefficient matrix consists of kappa coefficients & Sec.~\ref{subsubsec:simulation_for_analysis} \\
			$\mathcal{N}_{h}/\mathcal{N}_{l}$ & Category of agents, the average cooperation levels within their hyperedge is higher/lower than in their neighbors’ hyperedges & Eq.~\eqref{eq:N_hl} \\
			$\bar{f}_{s}^{h}/\bar{f}_{s}^{l}$ & Probability that initiators in $\mathcal{N}_{h}/\mathcal{N}_{l}$ are in state $s$  &  Eq.~\eqref{eq:fs2}\\
			$\bar{\phi}^{h}_{c|s}/\bar{\phi}^{l}_{c|s}$ & Cooperation preference for initiators in $\mathcal{N}_{h}/\mathcal{N}_{l}$ under the state $s$  	&  Eq.~\eqref{eq:phi2}		\\
			$[\kappa^{h}(s, s^{\prime})]/[\kappa^{l}(s, s^{\prime})]$  & Cohen’s kappa coefficient matrix in $\mathcal{N}_{h}/\mathcal{N}_{l}$ & Sec.~\ref{subsubsec:kappa2}\\
			$\hat{f}_{c}^{i}$ & Local cooperation levels within any hyperedge $e^{i}$ over a sliding time window with a specific scale & Eq.~\eqref{eq:hat_fc}
		\end{tabular}
	\end{ruledtabular}
\end{table*}


\bibliography{document}

\end{document}